\documentclass[
    10pt,                 %% Base font size
    twocolumn,            %% Two-column layout like final journal format
    preprintnumbers,      %% Show preprint numbers
    amsmath,              %% AMS math support
    amssymb,              %% Extra math symbols
    aps,                  %% Use APS formatting
    pre,                  %% Style for Physical Review E
    unsortedaddress       %% Authors listed in order of appearance (manual superscripts)
]{revtex4-2}

\usepackage{graphicx}       % For including images
\usepackage{array}          % Improved array and tabular environments
\usepackage{booktabs}       % Professional-quality tables
\usepackage{makecell}       % Multi-line and formatted table cells
\usepackage{bm}             % Bold math symbols
\usepackage[normalem]{ulem} % Underlining and strike-through text
\usepackage{xcolor}         % Colored text
\usepackage[colorlinks=true,linkcolor=black,urlcolor=blue,citecolor=red]{hyperref}
\usepackage[all]{hypcap}    % Fixes hyperlink anchors to floats
\newcommand{\supnum}[1]{\normalfont\textsuperscript{#1}} % Superscript numbers for affiliations
\usepackage{soul}

\begin{document}

\title{Multi-Scale Data Assimilation in Turbulent Models}

\author{
Francesco Fossella\supnum{1,2,3~\hyperlink{emailnote}{\textcolor{blue}{*}}},
Luca Biferale\supnum{2,3},
Alberto Carrassi\supnum{4},
Massimo Cencini\supnum{3,5},
Vikrant Gupta\supnum{6}
}

\affiliation{\vspace{0.1cm}
\mbox{\supnum{1}\textit{LTCI, Telecom Paris, 19 Place Marguerite Perey, 91120 Palaiseau, France.}}\\
\supnum{2}\textit{Department of Physics, University of Rome “Tor Vergata”, Via della Ricerca Scientifica 1, 00133 Rome, Italy.}\\
\mbox{\supnum{3}\textit{INFN “Tor Vergata”, Via della Ricerca Scientifica 1, 00133 Rome, Italy.}}\\
\supnum{4}\textit{Department of Physics and Astronomy, University of Bologna, Viale Carlo Berti Pichat 6/2, 40127 Bologna, Italy.}\\
\mbox{\supnum{5}\textit{CNR-ISC, Istituto dei Sistemi Complessi, Via dei Taurini 19, 00185 Rome, Italy.}}\\
\supnum{6}\textit{Department of Mechanical Engineering (Robotics), Guangdong Technion – Israel Institute of Technology, 241 Daxue Road, Shantou, Guangdong, 515063, China.}}

\begin{abstract}
We explore the potential of  Data-Assimilation (DA) within the multi-scale framework of a shell model of turbulence, with a focus on the Ensemble Kalman Filter (EnKF). The central objective is to understand how measuring mesoscales (i.e., inertial-range scales) enhances the prediction of both large-scale and small-scale intermittent variables, by systematically varying observation frequency and the set of measured scales. We demonstrate that measurements conducted at frequencies that exceed those of the observed scales enable full synchronization of larger scales, provided that at least two adjacent mesoscale are measured. In addition, we benchmark the EnKF against two other DA methods, namely Nudging and Ensemble 4D-Var. EnKF is clearly superior to the former, and comparable with the latter but achieving the result with a  lower computational complexity.
Moreover, our results underscore the need for a tailored, scale-aware inflation technique to stabilize
the assimilation process, preventing filter divergence and ensuring robust convergence.
\end{abstract}

\maketitle

%%%%%%%%%%%%% Section Introduction %%%%%%%%%%%%%
%%%%%%%%%%%%%%%%%%%%%%%%%%%%%%%%%%%%%%%%%%%%%%%%
\section{Introduction}
\label{sec:introduction}
The strongly nonlinear and multiscale character of turbulence severely limits the predictability horizon~\cite{lorenz_predictability_flow_1969,boffetta_musacchio_chaos_predictability_2017,aurell_boffetta_crisanti_paladin_vulpiani_predictability_1996,bandak_mailybaev_eyink_goldenfeld_spontaneous_stochasticity_2024}. Thus, even knowing the equations and providing measurements of some observables, the trajectories soon diverge from the prediction of the target evolution. A theoretical route to extend predictability is master–slave (or drive–response) synchronization: if a ``slave'' numerical model is continuously coupled to its ``master'' reference trajectory with a coupling strength above a well-defined threshold, the slave collapses onto the same chaotic attractor, thus eliminating the sensitivity to its own initial conditions \cite{pikovsky}. Complete chaotic synchronization, however, is feasible only when model and observations are error-free and every state variable is accessible, assumptions that are unattainable in realistic geophysical or engineering applications, where model errors, measurement noise, and sparse sampling are the rule. 

In order to overcome these limitations, the Data Assimilation (DA) framework was developed. Originally motivated by meteorological forecasting, DA provides systematic methods to combine model forecasts with sparse and noisy observations, improving the estimation of system states~\cite{talagrand, carrassi_2018, evensen_book}. In chaotic systems, DA offers a practical strategy to counteract the exponential divergence of trajectories by acting as an external forcing---analogous to a \emph{drag}---that competes with internal instabilities~\cite{carrassi_chaotic_dynamics_2022}. This competition naturally defines a synchronization threshold: When observational corrections outweigh the intrinsic divergence rate, synchronization can be achieved, and thus effective predictability.

This work mainly focuses on investigating the performance of a cornerstone DA technique, the Ensemble Kalman Filter (EnKF)~\cite{enkf_original, enkf_practical}, when applied to a shell model for turbulence~\cite{frisch, vulpiani, ditlevsen, sabra}. Our objective is to reconstruct the unobserved scales from sparse and noisy observations of the observed ones. Shell models emulate the turbulent energy cascade in a discretized Fourier space with logarithmically spaced wavenumbers, yielding several decades of scales and an equally wide spectrum of characteristic turnover times. Assimilating information measured at one scale to correct predictions at another is therefore an intrinsically multiscale challenge: updates must bridge slow, energy-containing modes and fast, highly intermittent ones. 

\begin{figure*}[]
    \centering
    \includegraphics[width=\textwidth]{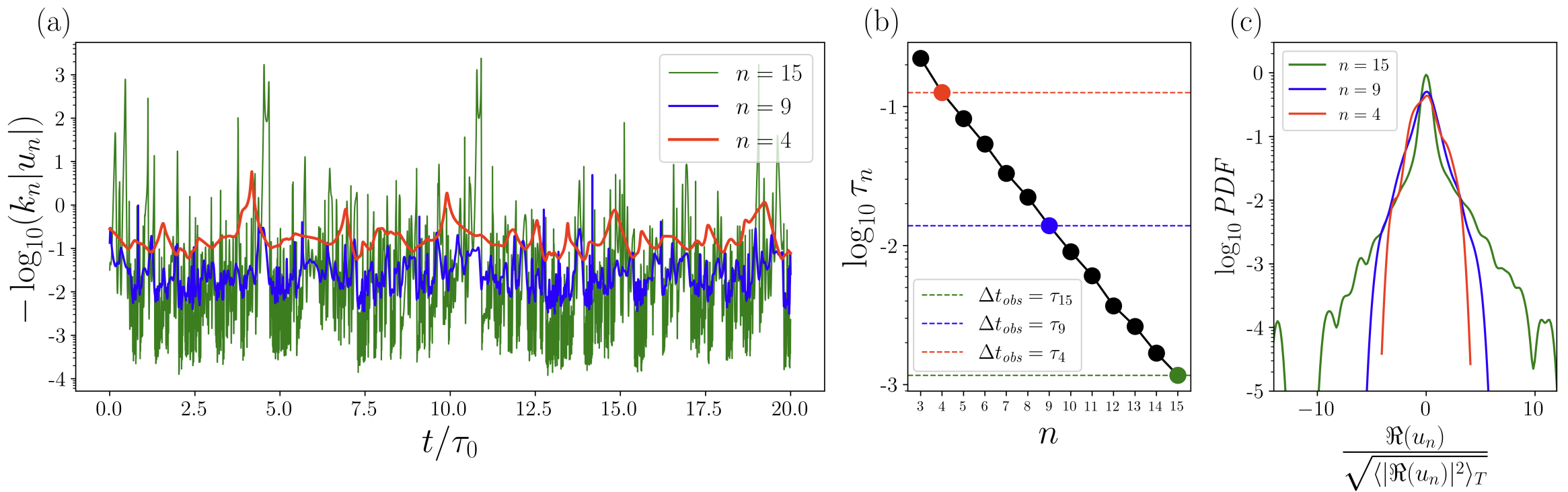}
    \caption{\textbf{Multi-scale character of the Sabra Model} (a) Instantaneous estimate of characteristic turnover times. 
        During one full oscillation of the slow shell $n = 4$, the fast shell $n = 15$---representative of the dissipative range---undergoes several hundred oscillations. (b) Quantitative visualization of the natural frequencies involved. 
        The observation intervals examined are highlighted: $\Delta t_{\text{obs}} = \tau_{15}$ yields quasi-continuous inertial-range measurements, $\Delta t_{\text{obs}} = \tau_{9}$ represents an intermediate regime, and $\Delta t_{\text{obs}} = \tau_{4}$ corresponds to discrete sampling. (c) Normalized probability-density functions of shell amplitudes. Slow shells display nearly Gaussian statistics, whereas fast shells exhibit heavy-tailed, intermittent behavior.}
\label{fig:turnover_time}
\end{figure*}

The EnKF advances an ensemble of model realizations and, at every assimilation instant, applies linear updates, provided the forecast and observation errors are Gaussian. Because the ensemble itself furnishes flow-dependent (prior) error covariances, the EnKF does not need to employ the tangent-linear and adjoint operators demanded by adjoint variational methods. However, in high-dimensional settings, computational constraints limit the ensemble size, introducing sampling noise that can erode covariance rank and trigger filter divergence~\cite{carrassi_2018,asch_data_assimilation_2016}. Practitioners therefore resort to inflation and covariance localization to maintain the need of ensemble spread and numerical stability \cite{gottwald_filter_divergence, sacher_bartello, inflation, raanes_adaptive_inflation_2019}. Despite these practical remedies, the EnKF and its extensions remain among the most widely used and readily implementable data assimilation methods for large-scale geophysical problems \cite{evensen_book, asch_data_assimilation_2016}. These limitations become particularly important in the multiscale, strongly intermittent dynamics of turbulence. The shell model considered here spans a few decades of scales and turnover times, and it faithfully reproduces the heavy-tailed, non-Gaussian statistics characteristic of small-scale turbulent fluctuations (see \hyperref[fig:turnover_time]{Fig.~\ref*{fig:turnover_time}(c)}). We adopt the \textit{stochastic} EnKF~\cite{enkf_practical} , which injects controlled Gaussian perturbations into the observations at every analysis step. This added noise mitigates the underestimation of the analysis uncertainty due to sampling error, enhances ensemble dispersion, and partly compensates for the non-Gaussian heavy tails generated by the fastest, most intermittent scales, thereby improving filter robustness in the highly nonlinear, multiscale regime addressed here. The present contribution therefore constitutes a first step toward systematically tailoring ensemble-based DA techniques to highly turbulent flows.

We shall benchmark the EnKF against two Data Assimilation (DA) methods at opposite ends of the DA complexity hierarchy. First, Nudging~\cite{nudging}, which can be viewed as a stripped-down single-member analogue of the EnKF. By adding continuous relaxation terms toward observations, it requires almost no statistical assumptions and is computationally inexpensive, but its strictly local corrections struggle to propagate information to unmeasured scales. The propagation of information from observations throughout the full domain is done solely via the model dynamics, and, as opposed to the EnKF it does not provide a quantification of the uncertainty in the state estimate. 

Second, we consider the so-called four‐dimensional variational method (4D-Var)~\cite{courtier_4dvar_1994}. 4D-Var seeks the model trajectory that minimizes a cost function over a finite assimilation window, thereby distributing corrections smoothly in time. The 4D-Var is formulated as a nonlinear optimization problem, as opposed to the linear-like update in the EnKF. Introducing time dependence in the forecast error covariance is very intricate in standard 4D-Var. This has been one of the main motivations driving the flourishing of hybrid ensemble variational methods~\cite{bannister_variational_review_2017,bocquet_ensemble_variational_2017}.  
In particular, when the background error covariance in the cost function is supplied fully or partly by an ensemble, the  formulation En4D-Var is obtained~\cite{ensemble_atmosperic}. The En4D-Var retains flow-dependent covariances but optimizes the entire trajectory within each window rather than issuing instantaneous updates. It is worth recalling that, in the linear case, 4D-Var and the Kalman filter are exactly equivalent and yield the same analysis at the end of the assimilation window, which makes their comparison particularly compelling in our strongly nonlinear model, where this equivalence no longer holds.

Data assimilation–based reconstruction from spectrally restricted observations has already been explored, by applying 4D-Var to a barotropic $\beta$-plane model to recover small-scale features~\cite{tanguay1995}, and by applying EnKF and nudging to the Sabra shell model~\cite{chen}. Futhermore, in the context of coupled DA, recent works have focused on adapting the EnKF to perform coupled atmospheric-ocean DA under various spatial and temporal scale separations and observational strength~\cite{tondeur2020,garcia2024}.
However, none of these studies have systematically varied the observational sparsity or the subset of observed scales, analyzing in detail the implications of DA on the energy cascade process.

In this work, we address these gaps by measuring different sets of shells at various observation sparseness, ranging from the slowest to the fastest time scales within the inertial range. This corresponds to passing from discrete to continuous sampling of
inertial range dynamics. This setup allows us to assess how the temporal resolution, relative to the local dynamical time scale, influences the assimilation process. We show that measuring at rates faster than the turnover time of the observed scales, and covering at least two adjacent shells, enables full synchronization of the larger scales. Under these conditions, the EnKF delivers superior performance, provided that appropriate scale-aware inflation is applied to mitigate sampling errors. We also discuss the limitations of Nudging in propagating corrections across scales and highlight the computational trade-offs associated with the use of EnKF and En4D-Var. We focus on a data assimilation setting in which both large and small scales are inferred from relatively sparse measurements at intermediate scales; although this is not the only possible configuration (for instance, one could directly observe the largest scales), it reflects the typical situation encountered in turbulent flows, where many degrees of freedom cooperate to sustain strongly out-of-equilibrium dynamics and no clear scale separation is available, in contrast to multiscale extensions of the Lorenz–96 model~\cite{lorenz96} where a slow–fast splitting is imposed a priori.

The remainder of the paper is organized as follows: In \hyperref[sec:shell_model]{Section~\ref*{sec:shell_model}}, we introduce the shell model. \hyperref[sec:enkf]{Section~\ref*{sec:enkf}} describes the EnKF algorithm, with special attention to the structure of the covariance and to the proposed inflation technique. \hyperref[sec:experimental_setup]{Section~\ref*{sec:experimental_setup}} details the experimental setup, while \hyperref[sec:enkf_results]{Section~\ref*{sec:enkf_results}} presents a comprehensive analysis of the EnKF results. Comparisons with Nudging and En4D-Var are discussed in \hyperref[sec:comparisons_nudging]{Section~\ref*{sec:comparisons_nudging}} and \hyperref[sec:comparisons_4dvar]{Section~\ref*{sec:comparisons_4dvar}}, respectively. Conclusions and future perspectives are summarized in \hyperref[sec:conclusion]{Section~\ref*{sec:conclusion}}, and details on the numerical integration of the model are given in \hyperref[app:rk4]{Appendix~\ref*{app:rk4}}.
%%%%%%%%%%%%% End Introduction Section %%%%%%%%%%%%%
%%%%%%%%%%%%%%%%%%%%%%%%%%%%%%%%%%%%%%%%%%%%%%%%%%%%

%%%%%%%%%%%%% Section Shell Model %%%%%%%%%%%%%
%%%%%%%%%%%%%%%%%%%%%%%%%%%%%%%%%%%%%%%%%%%%%%%
\section{Shell Models for Turbulence}
\label{sec:shell_model}
Proposed at the beginning of the 1970s~\cite{gledzer}, shell models provide significant insight into turbulent flow dynamics by mimicking the energy cascade at high Reynolds numbers~\cite{Biferale_shellmodel}. In these models, dynamics is discretized into $N$ consecutive scales, each represented by a wavenumber $k_n=2^n$ ($n = 0,1,\ldots,N-1$), commonly referred to as the ``shell'', with an associated complex velocity fluctuation $u_n\in\mathbb{C}$ evolving according to the laws:
\begin{equation}
    \label{eqn:shell_model}
    \left(\frac{{\rm d}}{{\rm d}t}+\nu k_n^2\right)u_n(t) = G_n[\bm{u}] + f_n(t)\ ,
\end{equation}
where $G_n[\bm{u}]$ represents the nonlinear coupling between the shells, and $f_n(t)$ is the external forcing. Essential to the model is the local, scale-by-scale, energy transfer between adjacent shells: energy is injected at large scales by $f_n$, it then cascades to smaller scales, where it is eventually dissipated by viscous effects, with $\nu$ being the viscosity parameter. 

The choice and design of the non-linear coupling, $G_n[\bm{u}]$, determines the qualitative behavior of the shell model in \hyperref[eqn:shell_model]{Eq.~(\ref*{eqn:shell_model})}. In the case of the Sabra shell model~\cite{sabra} the nonlinear coupling reads:
\begin{eqnarray}
    G_n[u] &=& i(ak_{n+1}u_{n+1}^*u_{n+2} + bk_n u_{n-1}^*u_{n+1} \nonumber \\
    &-& ck_{n-1}u_{n-1}u_{n-2})\, , \label{eqn:non_linear_coupling}
\end{eqnarray}
where $*$ denotes the complex conjugation.
The system of coupled differential equations is closed by imposing the boundary conditions $u_{-2} = u_{-1} = u_{N} = u_{N+1} = 0$.

We have chosen to work with this model because the two-point correlation is diagonal $\langle u_n u_{m}^* \rangle_T =0$ for $n\neq m$, and the only vanishing three-point correlations are the consecutive ones $\langle u_{n-1} u_{n}^* u_{n+1} \rangle_T$ (where $\langle\cdot\rangle_T$ denotes a time average taken in the statistically stationary regime). These features imply substantial analytical and numerical simplifications, enforce the locality of energy transfer, and arise from an inherent \textit{phase invariance} under the transformation $u_n\rightarrow u_n e^{i\theta}$, provided that
\begin{equation}
    \label{eqn:sabra_phase}
    \theta_{n+2}+\theta_{n+1}-\theta_{n}=0\ .
\end{equation}
In order to faithfully mimic a realistic fluid dynamical system, the inviscid and unforced case ($f_n=\nu=0$) must conserve the total energy 
$E(t)=\sum_{n=0}^N |u_n(t)|^2$, which imposes the constraint $a+b+c=0$ on the nonlinear coupling coefficients. The choice of $c$ is crucial in determining the physical dimensions of a second quadratic invariant postulated by the model. For $c<0$, the invariant is not positive definite, and selecting $c=-\frac{1}{2}$ assigns it the same physical dimensions as helicity $H(t)=\sum_{n=0}^N (-1)^n |u_n(t)|^2 k_n$, which is indeed conserved in the 3D Euler equations.

Other quantities of interest, which will be employed in subsequent analyses, are the structure functions. These are defined as 
\begin{equation}
    \label{eqn:structure_functions}
    S_p(k_n)= \left\langle|u_n|^p\right\rangle_T\ .
\end{equation}
In the inertial range they exhibit behavior $S_p(k_n)\propto k_n^{-\zeta(p)}$, with the exponents $\zeta(p)$ remarkably close to those measured in Navier-Stokes turbulence~\cite{sabra,Biferale_shellmodel}, but ``anomalous'' in the sense they deviate from the Kolmogorov prediction (1941)~\cite{frisch} $\zeta(p)=p/3$. The power law dependence highlights the multiscale character of the energy cascade and the presence of multiple characteristic timescales (\hyperref[fig:turnover_time]{Fig.~\ref*{fig:turnover_time}(a) and (b)}), which can be estimated as
\begin{equation}
\label{eqn:turnover-time}
\tau_n = \frac{1}{k_n\sqrt{\langle{|u_n|^2}\rangle_T}}\ .
\end{equation}
\noindent
Finally, the instantaneous energy flux up to the $n$-th scale is given by:
\begin{eqnarray}
    \label{eqn:flux}
    \Pi_n = \Im\left[a\, k_{n+1} u^*_n u^*_{n+1} u_{n+2} - c\, k_n u^*_{n-1} u^*_n u_{n+1}\right]\,.
\end{eqnarray}
With this convention for the flux definition, the combination of the two consecutive triads $u^*_n u^*_{n+1} u_{n+2}$ and $u^*_{n-1} u^*_n u_{n+1}$ represents rate of energy transferred within the spectral sphere up to the $n$-th shell.

We close this section mentioning that even though shell models may seem a simplification of 3D turbulent flows, in particular for the absence of a spatial structures, they preserve the multiscale character of turbulence with the presence of many characteristic times, which is one of the main difficulty when facing with predictability problems in turbulent flows. They represent a unique test bed where DA tools can be checked on fluid-like systems spanning many decades in frequencies and with strong non-Gaussians fluctuations (see \hyperref[fig:turnover_time]{Fig.~\ref*{fig:turnover_time}(c)}).
%%%%%%%%%%%%% End Shell Model %%%%%%%%%%%%%
%%%%%%%%%%%%%%%%%%%%%%%%%%%%%%%%%%%%%%%%%%%

%%%%%%%%%%%%% Section Ensemble Kalman Filter %%%%%%%%%%%%%
%%%%%%%%%%%%%%%%%%%%%%%%%%%%%%%%%%%%%%%%%%%%%%%%%
\section{Ensemble Kalman Filter for systems without circular symmetry }
\label{sec:enkf}
In addition to being inherently non-linear, shell models exhibit scale-dependent statistics, being approximately Gaussian at large scales but increasingly non-Gaussian with heavy tails at smaller ones. This variability requires a filtering method that is robust to deviations from Gaussianity. Among the available weakly non-linear ensemble Kalman filtering approaches, we adopt the stochastic variant in which each ensemble member assimilates a perturbed version of the observations~\cite{burgers_enkf_analysis}. In the stochastic EnKF, perturbations for the observations are drawn from the same Gaussian distribution assumed to describe the observational error. This choice mitigates the collapse of the analysis members onto the observations with consequent reduction of spread, the  underestimation of the posterior covariance~\cite{burgers_enkf_analysis,evensen_book},  and the impact of outliers~\cite{non-gaussianity}. We thus consider the stochastic EnKF to be particularly well-suited to the multi-scale statistical structure of shell models.

The Ensemble Kalman Filter operates by propagating an ensemble of model realizations, denoted $\{\hat{\bm{u}}^{(1)}, \hat{\bm{u}}^{(2)}, \ldots, \hat{\bm{u}}^{(L)}\} \in \mathbb{C}^{N \times L}$, where $N$ is the number of shells and $L$ is the number of ensemble members. In a Bayesian framework, these ensemble members represent samples from the prior distribution of the system state~\cite{bayesian_tutorial}. When new observations become available, the EnKF performs an analysis step that updates the prior ensemble to incorporate the observational information. This update uses empirical covariances computed from the ensemble to spread the influence of observed variables to unobserved ones. The result is a new set of ensemble members that approximate samples from the posterior distribution, i.e., the updated belief about the system state after assimilating the data. Although this update is formally linear, it is applied to ensemble members that have evolved under the nonlinear dynamics of the system. Consequently, the EnKF captures some nonlinear behavior and is thus considered a quasi-nonlinear or weakly nonlinear data assimilation method.

A pivotal concern lies in selecting a covariance structure that faithfully captures the behavior of the system. For complex-valued variables, the complex covariance matrix $\hat{\bm{\Sigma}}_{\mathbb{C}} = \langle(\hat{\bm{u}} - \langle \hat{\bm{u}} \rangle_L)(\hat{\bm{u}} - \langle \hat{\bm{u}} \rangle_L)^\dagger\rangle_L$ (where $\dagger$ denotes conjugate transpose and $\langle \cdot \rangle_L$ the ensemble average) encodes second-order statistics and the full signal power through its diagonal entries. However, $\hat{\bm{\Sigma}}_{\mathbb{C}}$ fully characterizes the statistics only under circular symmetry, i.e., invariance under global phase shifts $u_n \mapsto u_n e^{i\theta}$ for all $\theta \in [0, 2\pi]$~\cite{complex_kf}. When circular symmetry breaks, phase correlations emerge and complex covariance must be complemented by the pseudo-covariance matrix $\hat{\bm{\Sigma}}_p = \langle(\hat{\bm{u}} - \langle \hat{\bm{u}} \rangle_L)(\hat{\bm{u}} - \langle \hat{\bm{u}} \rangle_L)^{T}\rangle_L$, where ``T'' denotes matrix transposition. In shell models, the constraint (\hyperref[eqn:sabra_phase]{\ref*{eqn:sabra_phase}}) induces nontrivial phase correlations, preventing the pseudocovariance elements $(\hat{\Sigma}_p)_{nm}$ from vanishing for all $n, m$.

To correctly handle these correlations, we introduce the \emph{real extended covariance matrix} $\hat{\bm{\Sigma}} = \langle(\hat{\bm{U}} - \langle \hat{\bm{U}} \rangle_L)(\hat{\bm{U}} - \langle \hat{\bm{U}} \rangle_L)^{T}\rangle_L$, where $\hat{\bm{U}} = (\Re(\hat{\bm{u}}),\,\Im(\hat{\bm{u}}))\in\mathbb{R}^{2N}$. This representation retains both amplitude and phase information independently of any symmetry assumptions and takes the form:
\begin{equation}
\label{eqn:extended_real_covariance}
\hat{\bm{\Sigma}} = \begin{bmatrix}
\Re(\hat{\bm{\Sigma}}_{\mathbb{C}} + \hat{\bm{\Sigma}}_p) & -\Im(\hat{\bm{\Sigma}}_{\mathbb{C}} - \hat{\bm{\Sigma}}_p) \\
\Im(\hat{\bm{\Sigma}}_{\mathbb{C}} + \hat{\bm{\Sigma}}_p) & 
\ \ \Re(\hat{\bm{\Sigma}}_{\mathbb{C}} - \hat{\bm{\Sigma}}_p)
\end{bmatrix}\ ,
\end{equation}
which is symmetric due to the Hermitian property of $\hat{\bm{\Sigma}}_{\mathbb{C}}$, and positive definite as $\hat{\bm{U}}^\top \hat{\bm{\Sigma}} \hat{\bm{U}} \ge 0$.
Accordingly, both the system state and any operator must conform dimensionally to this $2N$-dimensional real representation.

\begin{figure*}[t]
    \centering
    \includegraphics[width=\textwidth]{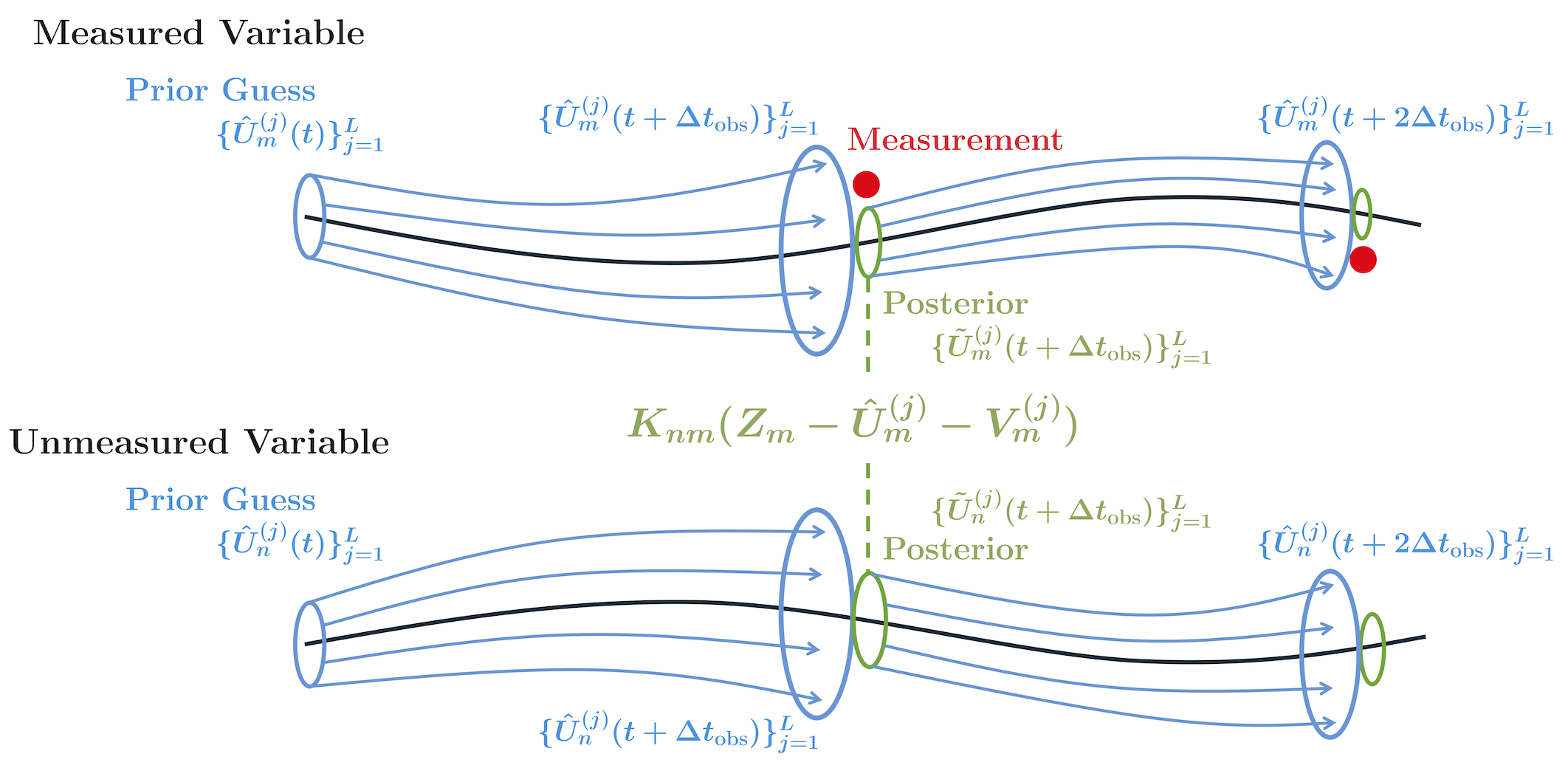}
    \caption{The ensemble (blue lines) starts as a prior distribution around the ground truth (black line) for both an observed variable $\hat{U}_m^{(j)}$ (red dots: measurements $Z_m$) and an unobserved one $\hat{U}_n^{(j)}$. Only two variables and $L=4$ ensemble members are shown. Due to chaotic divergence, the ensemble spread increases over time. At each assimilation step, measurements are incorporated to produce the posterior $\tilde{U}_m^{(j)}$ (green circles), which serves as the initial condition for the next cycle. The Kalman update propagates observational information to unmeasured variables, reducing their uncertainty as well.}
\label{fig:enkf_sketch}
\end{figure*}

The ground-truth state $\bm{U}(t)$ is approximated by an ensemble of prior estimates $\{\hat{\bm{U}}^{(1)}, \hat{\bm{U}}^{(2)}, \dots, \hat{\bm{U}}^{(L)}\} \subset \mathbb{R}^{2N}$ (the \textit{prior guess}, shown in \hyperref[fig:enkf_sketch]{Fig.~\ref*{fig:enkf_sketch}}), each evolving independently under the same nonlinear dynamics (perfect-model scenario), given by \hyperref[eqn:shell_model]{Eq.~(\ref*{eqn:shell_model})} expressed in real-valued form. The ensemble is advanced to the next observation time, $t_{\text{obs}} = t + \Delta t_{\textit{obs}}$, when a new set of measurements $\bm{Z}(t_{\text{obs}}) \in \mathbb{R}^{2M}$ (with $M \leq N$) of the ground-truth state $\bm{U}(t_{\text{obs}})$ becomes available (indicated as a \textit{red dot} in \hyperref[fig:enkf_sketch]{Fig.~\ref*{fig:enkf_sketch}}):
\begin{equation}
    \label{eqn:measurements}
    \bm{Z}(t_{\text{obs}}) = \bm{H}\bm{U}(t_{\text{obs}}) + \bm{\xi}(t_{\text{obs}})\,,
    \quad \bm{\xi}(t_{\text{obs}}) \sim \mathcal{N}(\bm{0}, \bm{R})\,.
\end{equation}
The linear observation operator $\bm{H} \in \mathbb{R}^{2M \times 2N}$ is a selection matrix containing only 0s and 1s, used to extract the measured variables (real or imaginary parts of the shells), and $\bm{\xi}(t_{\text{obs}})$ denotes the measurement noise. The observation error covariance matrix \( \mathbf{R}\in \mathbb{R}^{2M \times 2M} \) is generally time-dependent. However, in our experiment, it is assumed to be diagonal and time independent (see \hyperref[measurement_errors_setup]{Measurement Errors} in \hyperref[sec:experimental_setup]{Sec.~\ref{sec:experimental_setup}}).

\noindent
To assimilate the measurements, we apply the stochastic EnKF update~\cite{leeuwen}, where the observations are perturbed by independent realizations $\bm{V}^{(j)}(t_{\text{obs}}) \sim \mathcal{N}(\bm{0}, \bm{R})$. The posterior ensemble is updated according to:
\begin{eqnarray}
\label{eqn:ensemble_update}
\tilde{\bm{U}}^{(j)}(t_{\text{obs}}) 
&=& \hat{\bm{U}}^{(j)}(t_{\text{obs}}) \nonumber \\
&&\hspace{-5em} +\!\bm{K}(t_{\text{obs}})\! \left[ \bm{Z}(t_{\text{obs}})\! 
-\!\bm{H} \hat{\bm{U}}^{(j)}(t_{\text{obs}}) 
\!-\! \bm{V}^{(j)}(t_{\text{obs}}) \right]\,,
\end{eqnarray}
with $j = 1, \ldots, L$. This analysis step provides the updated posterior variable (denoted by \textasciitilde) as a linear combination of the prior (denoted by \textasciicircum) and the measurements \( \bm{Z} \) of the ground truth. The Kalman gain \( \bm{K}(t_{\text{obs}}) \in \mathbb{R}^{2N \times 2M} \) is computed as~\cite{evensen_book}:
\begin{equation}
    \label{eqn:kalman_gain}
    \bm{K}(t_{\text{obs}}) = \hat{\bm{\Sigma}}(t_{\text{obs}})\,\bm{H}^{T} \left(\bm{H} \hat{\bm{\Sigma}}(t_{\text{obs}}) \bm{H}^{T} + \bm{R}\right)^{-1}\ ,
\end{equation}
and the prior covariance matrix $\hat{\bm{\Sigma}}(t_{\text{obs}})$ is estimated from the ensemble via:
\begin{equation}
    \label{eqn:prior_covariance}
    \hat{\bm{\Sigma}}(t_{\text{obs}}) = \frac{1}{L-1} \sum_{j=1}^L \left[\hat{\bm{U}}^{(j)} - \hat{\bm{\mu}}\right] \left[\hat{\bm{U}}^{(j)} - \hat{\bm{\mu}}\right]^{T}\ ,
    \quad
\end{equation}
where $\hat{\bm{\mu}} = \frac{1}{L} \sum_{j=1}^L \hat{\bm{U}}^{(j)}$ is the prior ensemble mean.

Each element $K_{nm}$ of the Kalman gain quantifies the impact of the $m$-th observed component on the correction of the $n$-th state variable. Specifically, the correction applied to $\hat{U}^{(j)}_n$ is given by $K_{nm}\left[\bm{Z} - \bm{H}\hat{\bm{U}}^{(j)} - \bm{V}^{(j)}\right]_m$, where the Einstein summation convention is implied over repeated indices. This formulation ensures that measurement information is spread across all components proportionally to their statistical correlations.

The posterior ensemble moves closer to the observations when $\bm{R}$ is small and remains near the prior when $\hat{\bm{\Sigma}}(t_{\text{obs}})$ is small. For Gaussian errors and linear observations of the forecast model, the Kalman gain optimally balances these contributions by minimizing the trace of posterior covariance $\tilde{\bm{\Sigma}}$, satisfying $\partial\,\mathrm{tr}(\tilde{\bm{\Sigma}})/\partial \bm{K} = 0$~\cite{asch_data_assimilation_2016}. Stochastic perturbations $\bm{V}^{(j)}$ ensure that the updated ensemble remains a consistent sample of the filtering distribution: specifically, they avoid underestimating the ``theoretical'' posterior covariance~\cite{burgers_enkf_analysis}. 
The posterior covariance is then obtained recursively as in the Kalman filter update according to 
$\tilde{\bm{\Sigma}}(t_{\text{obs}}) = (\bm{I} - \bm{K}\bm{H}) \hat{\bm{\Sigma}}(t_{\text{obs}})$, from where we see that its determinant is bound to be smaller or equal to one, reflecting a reduction of uncertainty. The posterior ensemble is obtained directly, member-by-member, by executing \hyperref[eqn:ensemble_update]{Eq.~(\ref*{eqn:ensemble_update})} (\textit{green circles} in \hyperref[fig:enkf_sketch]{Fig.~\ref*{fig:enkf_sketch}}). The EnKF thus solves sequentially, likewise the Kalman filter~\cite{KF}, the state-estimation problem, by alternating the forecast (prior) and the analysis (posterior) steps, until the onset of the \emph{saturation regime}, where the ensemble mean has likely converged near to the true state. A schematic illustration of the functioning of the EnKF is provided in \hyperref[fig:enkf_sketch]{Fig.~\ref*{fig:enkf_sketch}}. 

\subsection{Scale Aware Inflation Strategy} 
\label{sub:inflation}
As a Monte Carlo-based method, EnKF is inherently prone to instability issues due to sampling errors. These arise from the inevitable limited number of ensemble members in comparison to what is needed to properly describe the system’s effective degrees of freedom~\cite{carrassi_2018}. For Gaussian and linear systems, the skill of stochastic EnKF monotonically converges to the ideal KF in the limit of an infinitely large ensemble. The minimum number of members required for satisfactory performance and computational affordability depends on the model-and- observational scenarios, as well as on the type of EnKF. For chaotic dynamics, in principle, there is a relation with the number of unstable modes of dynamics~\cite{carrassi_chaotic_dynamics_2022}. When the ensemble size is small, the estimated covariance matrix tends to contain spurious correlations, typically manifesting as underestimated variances and overestimated cross-correlations at large distance. While the latter are usually understood in physical space, the same is often the case also in spectral domain, whereby the actual correlation between scales/frequencies decreases with the frequency gap, while the ensemble-based correlations do not decrease similarly fast or not at all. These artificial, non-physical correlations pose a serious problem: when a new observation is assimilated, the update step may inadvertently modify variables that should remain unaffected, excessively reducing their associated error variance. In the next assimilation cycle, the ensemble spread becomes so narrow that even small, incorrect adjustments can propagate across unrelated variables. The EnKF assigns insufficient weight to incoming observations, leading to filter divergence, which is particularly problematic in chaotic systems~\cite{grudzien_covariance_inflation}. The resulting error perturbations increase the stiffness of the system and can ultimately cause catastrophic divergence, a finite-time numerical blow-up of the forecast~\cite{gottwald_filter_divergence}. \noindent The indisputable success of EnKF methods in high-dimension chaotic systems is also related to the development of a plethora of approaches, and ad-hoc fixes, to counteract the effect of the sampling error. These efforts can all be framed under two algorithmic solutions: {\it localization} and {\it inflation}~\cite{carrassi_2018,evensen_book,asch_data_assimilation_2016}. Localization consists in restricting the effect of the analysis update to the vicinity (or local domain) of each observation and is a widely used and often indispensable technique in medium- to large-scale systems. The size of the localization domain is usually chosen based on practical arguments and may require numerical tuning, although recent studies have introduced scale-dependent localization approaches~\cite{pasmans_tailoring_da_2024,menetrier_linear_filtering_2015}. In the shell model, the energy cascade is inherently local, with interactions occurring primarily between neighboring scales. This makes it natural to impose a tridiagonal structure on the covariance matrix as a form of spectral localization. However, our experiments reveal that such localization---either in the form of a tridiagonal approximation or of a stricter spectral correlation cutoff---while successful in eliminating spurious correlations due to sampling error, leads to physically imbalanced analyses causing larger forecast errors. We therefore omit using localization and instead shall demonstrate that an appropriate inflation strategy will suffice. We introduce a novel scale-aware inflation approach that mitigates the effects of undersampling without sacrificing physical consistency. Inflation consists of artificially increasing the ensemble-based error covariance~\cite{inflation}, and is usually performed in two alternative ways, {\it additive} or {\it multiplicative}~\cite{carrassi_2018}. It can be applied to either the forecast (prior) or the analysis (posterior) error covariance, with different effects. In additive inflation, a random term is added to each ensemble member to maintain the diversity and the spread of the ensemble. In multiplicative inflation, the ensemble-based error covariance, either prior or posterior, is inflated by a scalar factor. This inflates the ensemble variance by rescaling each posterior ensemble member around the mean as: \begin{equation} \label{eqn:inflation} g_n(t) \tilde{U}_n+(1-g_n(t))\tilde{\mu}_n\to \tilde{U}_n\ , \end{equation} where $\tilde{\mu}_n$ is the posterior mean and $g_n(t)\ge 1$ is the inflation coefficient. This transformation increases the dispersion of the ensemble around the mean and rescales the covariance element $\tilde{\Sigma}_{nm}$ to $g_n g_m \tilde{\Sigma}_{nm}$. Sacher and Bartello~\cite{sacher_bartello} showed that the sampling error in the Kalman gain estimate tends to scale directly proportional to the magnitude of the gain itself. This implies that larger corrections from observations require stronger inflation to maintain stability. It is therefore desirable to inflate the ensemble more in regions where observations exert a stronger influence. Motivated by this, we propose a scale-aware inflation scheme, where the spread reduction is compensated by: \begin{equation} \label{eqn:inflation_coeff} g_n(t)=\textit{max}\left(1,1+\lambda\frac{\hat{\Sigma}_{nn}-\tilde{\Sigma}_{nn}}{\hat{\Sigma}_{nn}}\right)\ , \end{equation} with $\hat{\Sigma}_{nn}$ and $\tilde{\Sigma}_{nn}$ denoting the prior and posterior variances, respectively, and $\lambda\in{\mathbb R}^+$ the inflation strength, which is a hyperparameter to be tuned. This formulation is inspired by the original work of Zhang et al.~\cite{RTP}, which used a similar expression but normalized by the posterior variance $\tilde{\Sigma}_{nn}$ instead of ${\hat{\Sigma}_{nn}}$. Since, statistically, $\langle \hat{\Sigma}_{nn} \rangle_T \geq \langle \tilde{\Sigma}_{nn} \rangle_T$, our version yields a smaller inflation coefficient for the same spread variation and $\lambda$, resulting in a less aggressive inflation. The choice of the parameter $\lambda$ is dependent on the system and must be carefully tuned to ensure effective assimilation. Although our inflation strategy is scale-sensitive, tuning $\lambda$ adds computational cost. A detailed analysis of its effect on analysis statistics, filter stability, and comparison with other techniques lies beyond the scope of this work. However, even with extensive adjustment of $\lambda$, the proposed inflation strategy turns out to be very successful. It has been effective in stabilizing the assimilation process and in substantially reducing the minimum ensemble size needed to achieve satisfactorily performance and avoid filter divergence. Further and dedicated developments of the proposed scale-aware inflation method will be the content of follow-up works. 

Numerical details on inflation prefactor $\lambda$ used are provided in \hyperref[app:inflation]{Appendix~\ref{app:inflation}}.
%%%%%%%%%%%%% End Section EnKF %%%%%%%%%%%%%%
%%%%%%%%%%%%%%%%%%%%%%%%%%%%%%%%%%%%%%%%%

%%%%%%%%%%%%% Section Experimental Setup %%%%%%%%%%%%%
%%%%%%%%%%%%%%%%%%%%%%%%%%%%%%%%%%%%%%%%%%%%%%%%%%%
\section{Numerical  Setup}
\label{sec:experimental_setup}
We replicate numerically the experimental scenario in which measurements are only available for the \emph{mesoscale} ({\it i.e.}, inertial range scales), which in principle are more accessible to mesuraments. The goal is to study the feasibility and capability of EnKF to infer simultaneously, based solely on these mesoscale measurements, the large- and the small-intermittent scale variables. The latter are usually very difficult or even impossible to measure in real applications. 

In this study, we work under the \textit{perfect-model} setting, implying that the Sabra shell model used as ground truth, and from which synthetic data are extracted, is the same as the model used to assimilate such data. The prediction error arises only from imperfect initial conditions. The model parameters are set to $(a, b, c) = (1, -0.5, -0.5)$, with the number of shells $N = 20$, viscosity $\nu = 10^{-6}$, and time step $dt = 10^{-5}$. A constant forcing term, $f_n = (1 + i)\delta_{n0}$ ($\delta_{n0}$ is the Kronecker symbol), is applied exclusively to the first shell. The selection of these parameters ensures the stability of the numerical solver as detailed in \hyperref[app:rk4]{Appendix~\ref*{app:rk4}}.

Since both the ensemble and the ground-truth solution evolve under the same governing equations, the setup constitutes a \emph{twin experiment} in DA jargon. Initial conditions' errors are (generally) amplified because of the chaotic nature of the underlying dynamics. To ensure that the initial conditions for the ground truth, $\bm{u}$, are sampled from the system attractor, {\it that is} they are statistically consistent with the system dynamics, a preliminary spin-up integration is carried out over a sufficiently long time interval, over many turnover times $\tau_0$ of the first-forced shell (\hyperref[eqn:turnover-time]{Eq.~(\ref*{eqn:turnover-time}})). This guarantees that we assimilate data from a regime characterized by well-established statistical properties. 
For EnKF, a larger initial spread allows the ensemble to explore a wider region of the state space, making it more compatible with the first observation to be assimilated and thereby improving filter stability. This also helps mitigating EnKF-specific instabilities and can reduce the need for artificial inflation. To achieve this, the ensemble is freely evolved for $10\tau_0$ before assimilation begins. During this spin-up phase, chaotic dynamics naturally increases the ensemble spread, although it may also reduce the individual member diversity by making them align along the most unstable modes. Since chaotic systems rapidly lose memory of their initial conditions, the specific form of the initialization becomes irrelevant and can therefore be chosen as:
\begin{equation}
\label{eqn:initial_guess_shell_condition_enkf}
\hat{u}_n^{(j)}(t_0) = \sqrt{|u_n(t_0)|^2} \left[ \cos\left(\phi_n^{(j)}\right) + i \sin\left(\phi_n^{(j)}\right) \right]\ ,
\end{equation}
where the phases $\phi_n^{(j)}$ are independent random variables uniformly chosen in $[0,2\pi]$.
This approach ensures dynamical balance within the ensemble and promotes the emergence of physically meaningful multivariate correlations~\cite{enkf_practical}. \\
\noindent {\textbf{Measurement Errors}}
\label{measurement_errors_setup}
Both the real and imaginary parts of each measured shell are observed simultaneously, with independent Gaussian noise drawn from $\mathcal{N}(\mathbf{0}, \mathbf{R})$, where the covariance matrix $\mathbf{R}$ is defined as:
\begin{equation}
    \label{eqn:meas_cov_matrix}
    R_{mn} = \delta_{mn}(0.05)^2\langle |u_{\lfloor m/2 \rfloor}|^2 \rangle_T\ ,
\end{equation}
with $m, n = 0, \ldots, 2N-1$ indexing the measured components of the extended state vector, and $\lfloor \cdot \rfloor$ denoting the floor function. This formulation ensures that both the real and imaginary components are measured with the same error variance, proportional to the energy of the corresponding shell. Notice that the error, chosen to be $5\%$ of the standard deviation of the measured variables, is relatively large when compared to the large-scale variability, where the statistics are approximately Gaussian, and very small at small scales,  where energy fluctuations for each shell can easily reach 8-10 times the standard deviation (see \hyperref[fig:turnover_time]{Fig.~\ref*{fig:turnover_time}(c)}).

The diagonal structure of $\mathbf{R}$ reflects the assumption that measurement errors across different shells are independent, implying that the observation process does not introduce cross-scale error correlations. The same structure is used for the perturbation noise required by the EnKF procedure.
\\
\noindent
{\textbf{Evaluation Metrics}}
\label{evaluation_metrics}
To evaluate the effectiveness of the DA method in recovering the ground-truth signal, we compute the mean squared errors (MSEs) averaged in time and, when applicable, over the ensemble. To this end, we define the temporal average over $N_T$ analysis steps as $\langle \cdot \rangle_T = \frac{1}{N_T} \sum_{t_k=1}^{N_T} (\cdot)$, and the ensemble average over $L$ members as $\langle \cdot \rangle_L = \frac{1}{L} \sum_{j=1}^{L} (\cdot)$. Their combination yields the joint average $\langle \cdot \rangle_{T,L} = \langle \langle \cdot \rangle_L \rangle_T$, used to define the MSE at the $n$-th shell:
\begin{equation}
    \label{eqn:error_usual}
    \mathcal{E}_n = \langle |u_n - \tilde{u}_n|^2 \rangle_{T,L}\ ,
\end{equation}
where $\tilde{u}_n$ is the assimilated variable (posterior). In the case of a single estimate, such as for the ground truth or nudging, the ensemble average is dropped, and only the time average is applied. The total number of time steps $N_T$ depends on the observation frequency, as detailed in the results section.

A second diagnostic concerns the energy flux (\hyperref[eqn:flux]{Eq.~(\ref*{eqn:flux})}), based on the triadic interaction $u_{n-1} u_n u^*_{n+1}$, with the error defined as:
\begin{equation}
    \label{eqn:error_flux}
    \mathcal{E}^{\Pi}_n = \langle |u_{n-1} u_n u^*_{n+1} - \tilde{u}_{n-1} \tilde{u}_n \tilde{u}^*_{n+1}|^2 \rangle_{T,L}\ .
\end{equation}
We also define the average ground-truth and ensemble power spectra energies as:
\begin{equation}
    \label{eqn:mean_energies}
    E_n=\langle |u_n|^2 \rangle_T\ , \qquad \tilde{E}_n=\langle |\tilde{u}_n|^2 \rangle_{T,L} 
\end{equation} 
and the corresponding triadic terms entering in the energy flux  as $\Delta_n=\langle |u_{n-1} u_n u^*_{n+1}|^2 \rangle_T$, $\tilde{\Delta}_n=\langle |\tilde{u}_{n-1} \tilde{u}_n \tilde{u}^*_{n+1}|^2 \rangle_{T,L}$, allowing us to normalize the MSEs as $\mathcal{E}_n/\sqrt{\tilde{E}_n E_n}$ and $\mathcal{E}^{\Pi}_n/\sqrt{\tilde{\Delta}_n \Delta_n}$.\\
Since the result may vary with different ground-truth realizations, we performed $N_\textit{exp}=16$ independent runs and reported the final errors as the average between maximum and minimum between experiments. For example, the normalized 
MSE is given by:
\begin{equation}
\label{eqn:error_bars}
\begin{aligned}
&\frac{|\max_{N_\textit{exp}}(\mathcal{E}_n/\sqrt{\tilde{E}_nE_n}) + \min_{N_\textit{exp}}(\mathcal{E}_n/\sqrt{\tilde{E}_nE_n})|}{2}\\
\pm&\frac{|\max_{N_\textit{exp}}(\mathcal{E}_n/\sqrt{\tilde{E}_nE_n}) - \min_{N_\textit{exp}}(\mathcal{E}_n/\sqrt{\tilde{E}_nE_n})|}{2}\,,
\end{aligned}
\end{equation}
where the second term represents the error bar, capturing the variability between experiments. With a slight abuse of notation, we shall continue to denote all diagnostic quantities introduced above---including the energies \( E_n \), \( \tilde{E}_n \), the errors \( \mathcal{E}_n \), \( \mathcal{E}_n^\Pi \), and their normalized counterparts---by the same symbols, even though they are implicitly averaged over the set of independent experiments, as just described.

%%%%%%%%%%%%% End Section Experimental Setup %%%%%%%%%%%%%
%%%%%%%%%%%%%%%%%%%%%%%%%%%%%%%%%%%%%%%%%%%%%%%%%%%
\begin{figure*}[]
    \centering
    \includegraphics[width=\textwidth]{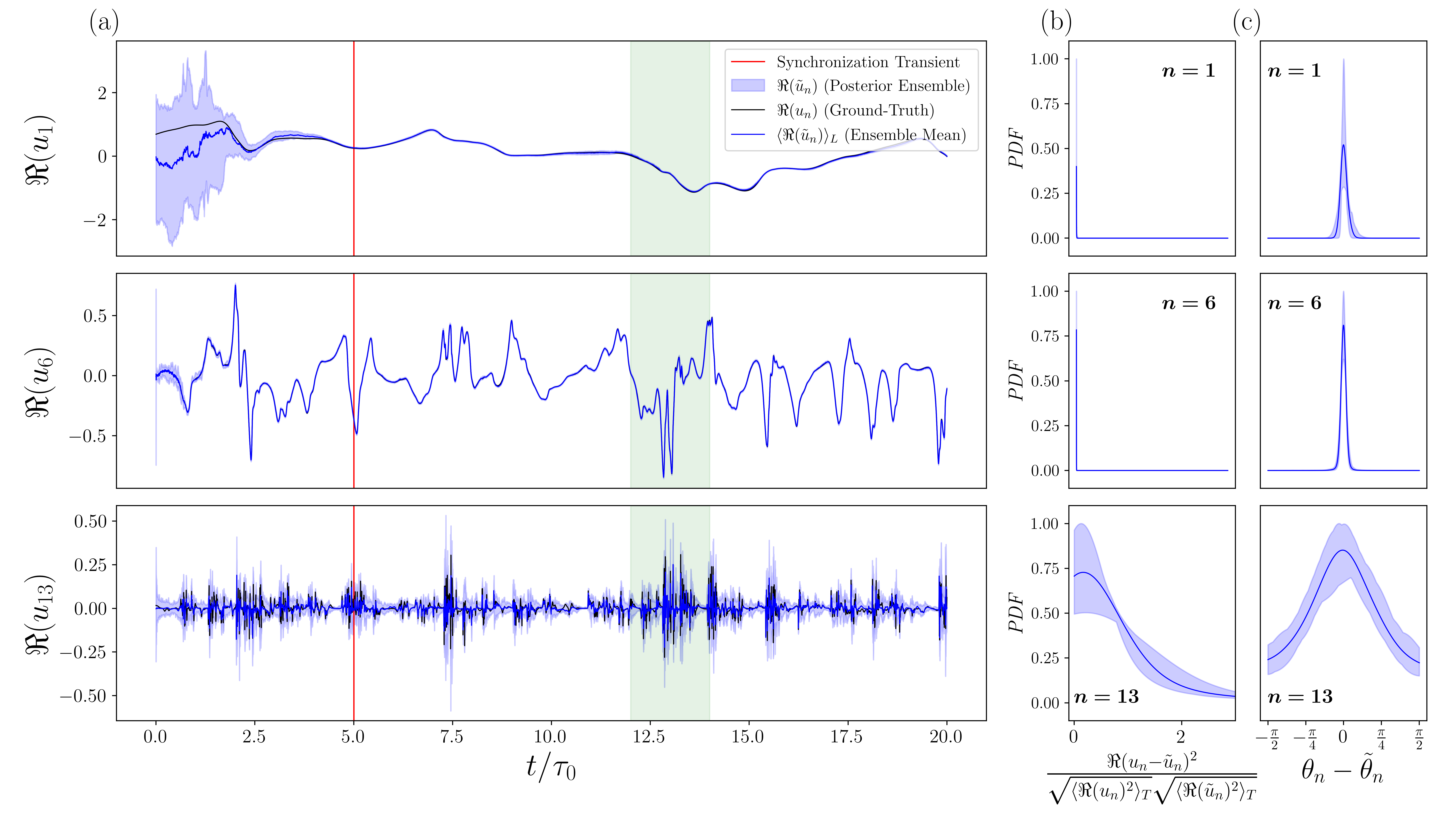}
    \caption{(a) Real parts of the velocity components for shells $n=1$, $6$, and $13$ from an experiment where $u_6$, $u_7$, and $u_8$ are measured with observation interval $\Delta t_{\text{obs}} = \tau_{15} = 0.002\tau_0$. The figure displays the full ensemble (light blue), its mean (solid blue line), and the ground truth (solid black line), over a total experiment duration of $20\tau_0$. (b) Probability density function (PDF) of the normalized real-part error, $\frac{\Re(u_n-\tilde{u}_n)^2}{\sqrt{\langle\Re(u_n)^2\rangle_T}\sqrt{\langle\Re(\tilde{u}_n)^2\rangle_T}}$, and (c) PDF of the phase difference, $\theta_n - \tilde{\theta}_n$, shown for all ensemble members (shaded area) and their mean (blue line) for the selected shells. The PDFs in (b) and (c) are computed by collecting data only after the transient phase---i.e., after saturation is reached (as indicated by the red vertical line in panel (a))---and up to the end of the $20\tau_0$ experiment. The light-green shaded zone highlights the time window $12\leq t/\tau_0\leq 14$ detailed in \hyperref[fig:all_together]{Fig.~\ref*{fig:all_together}(a) and (b)}.}
\label{fig:dinamic_evol_enkf}
\end{figure*}

\begin{figure*}[]
    \centering
    {\includegraphics[width=\textwidth]{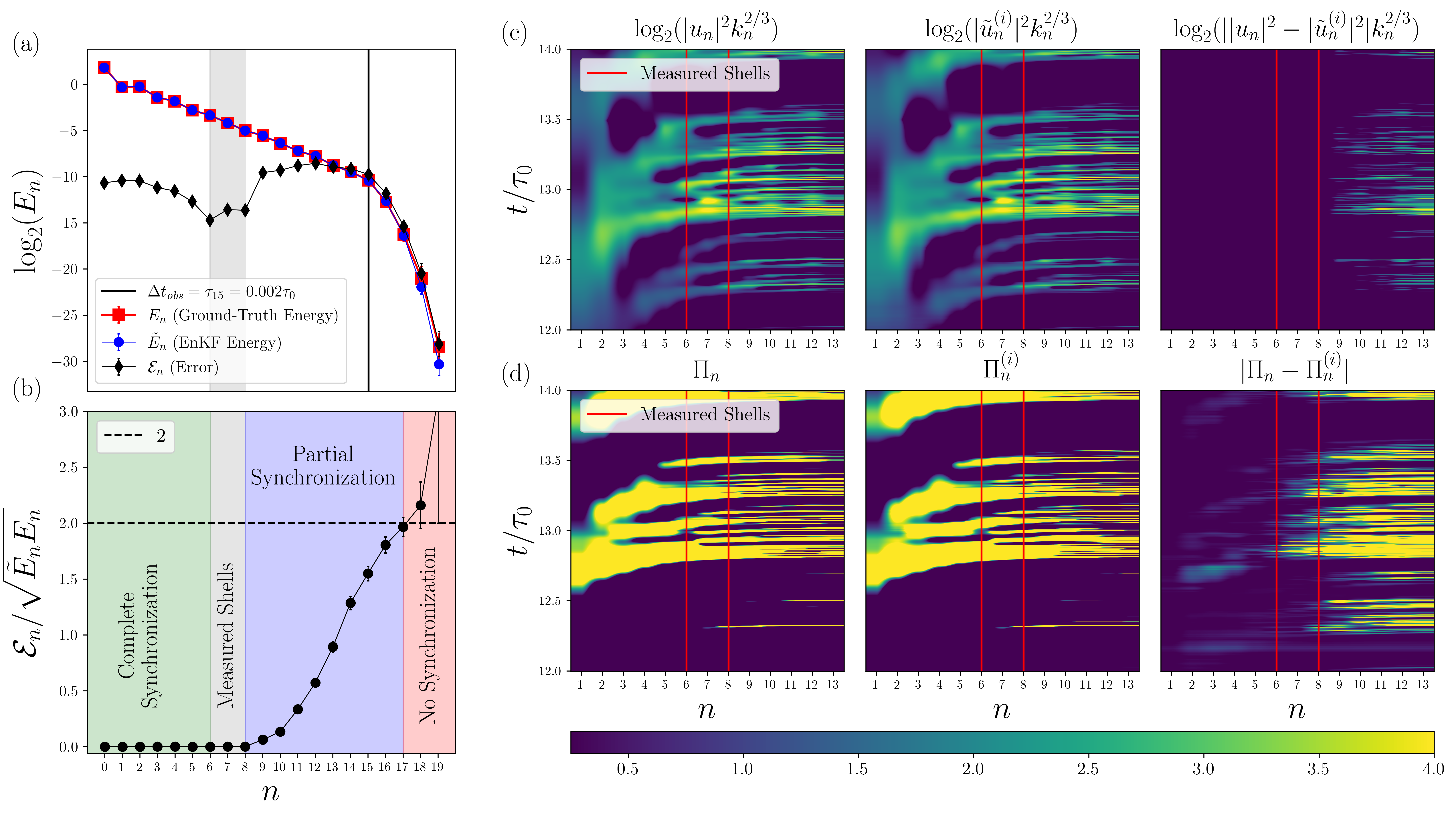}}
    \caption{(a) Ground truth energy spectrum $E_n = \langle |u_n|^2 \rangle_T$, EnKF estimate $\tilde{E}_n = \langle |\tilde{u}_n|^2 \rangle_{T,L}$, and $\ell^2$-error $\mathcal{E}_n = \langle |u_n - \tilde{u}_n|^2 \rangle_{T,L}$. The grey shaded area highlights the observed shells, while the vertical solid line marks the scale for which the turnover time matches the observation interval, $\tau_{15} = \Delta t_{\textit{obs}}$. (b) Normalized error $\mathcal{E}_n / \sqrt{E_n \tilde{E}_n}$, with the dashed line indicating the scale beyond which statistical consistency ($E_n = \tilde{E}_n$) is maintained, but the EnKF no longer achieves effective reconstruction. All quantities in panels (a) and (b), including error bars, are averaged over $N_{\text{exp}} = 16$ independent experiments, as described in \hyperref[sec:experimental_setup]{Section~\ref*{sec:experimental_setup}}. (c) Normalized energy evolution for inertial-range scales during the stationary synchronization regime ($12 \leq t / \tau_0 \leq 14$; see light-green shaded region in \hyperref[fig:dinamic_evol_enkf]{Fig.~\ref*{fig:dinamic_evol_enkf}(a)}). Shown are the ground truth values $|u_n|^2 k_n^{2/3}$ (left), one randomly selected posterior ensemble member $|\tilde{u}_n|^2 k_n^{2/3}$ (middle), and their difference (right). Red vertical lines indicate the measured shells ($n=6$, $7$, and $8$). (d) Cumulative instantaneous energy flux evolution over the same stationary synchronization regime, shown for the ground truth (left), one randomly selected posterior ensemble member (middle), and their difference (right).}
    \label{fig:all_together}
\end{figure*}

\begin{figure}[]
    \centering
    \includegraphics[width=0.45\textwidth]{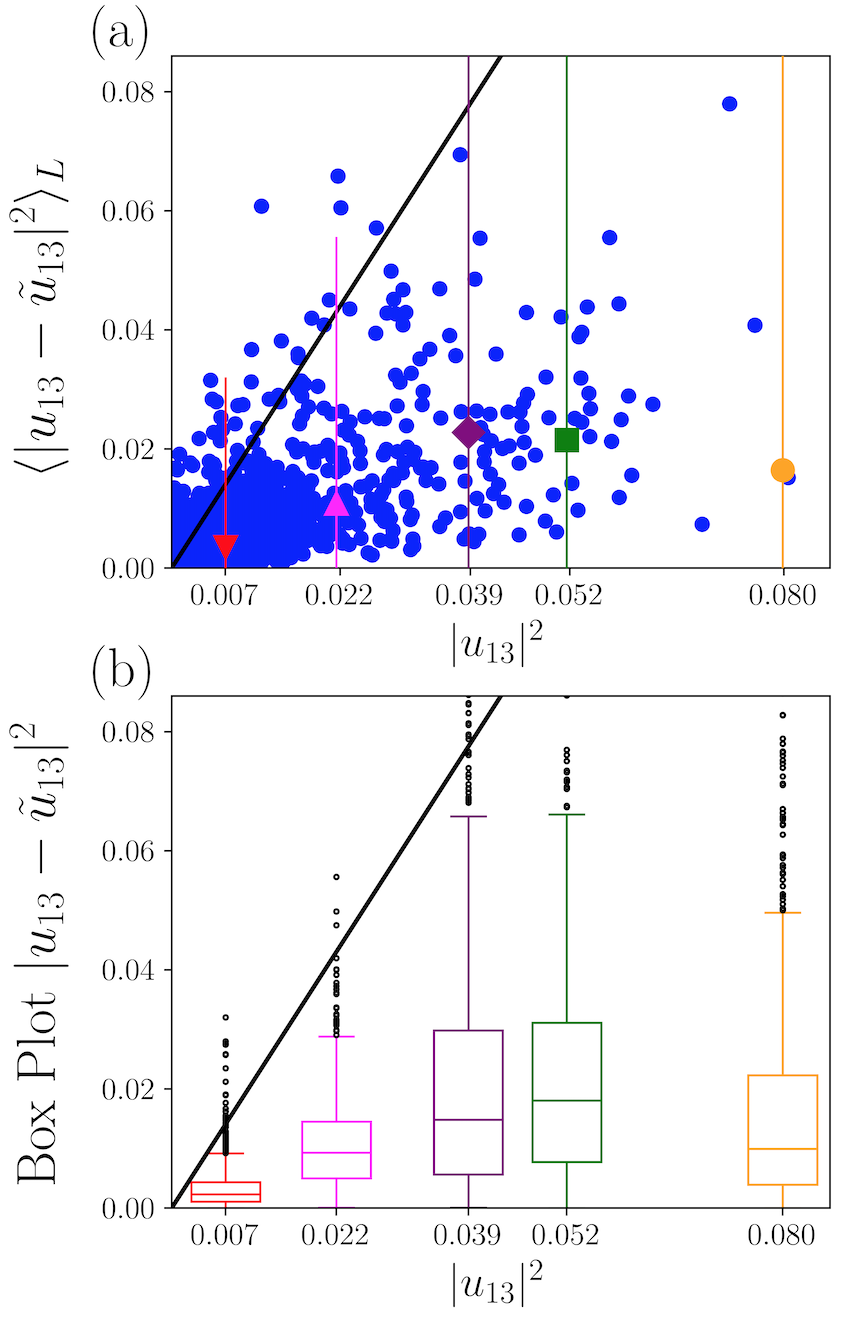}
    \caption{(a) Scatter plot of the mean errors $\langle|u_{13}-\tilde{u}_{13}|^2\rangle_L$ versus the energy $|u_{13}|^2$ at scale $n=13$, including the statistical independence baseline given by $\langle|u_{13}-\tilde{u}_{13}|^2\rangle_L=2|u_{13}|^2$, and five highlighted specific cases. Colored vertical lines through each point illustrate the full distribution of ensemble errors $|u_n-\tilde{u}_n^{(j)}|^2$, providing insight into variability around the mean value. (b) Box plots for the highlighted points in Panel~(a). In each box plot, the box spans the interquartile range (from the 25th to the 75th percentile), with the horizontal line inside indicating the median. The vertical whiskers extend to the next percentiles (representing the 10th and 90th percentiles, or the most extreme non-outlier values), while black dots denote outliers. The results shown in these panels were collected from an experiment lasting $20\tau_0$.}
    \label{fig:extreme_events}
\end{figure}

%%%%%%%%%%% Section Ensemble Kalman Filter Results %%%%%%%%%%%%%%
\section{Ensemble Kalman Filter Results}
\label{sec:enkf_results} 
The shell model is a multi-scale system in which progressively smaller and faster scales are characterized by their respective turnover times $\tau_n$. This naturally raises the question of whether there exists an optimal assimilation window, i.e., the time interval between consecutive measurements $\Delta t_{\textit{obs}}$, such that setting $\Delta t_{\textit{obs}} = \tau_n$ minimizes reconstruction errors not only on the observed scales but also on neighboring ones, while preserving the statistical properties of the system. A second key question is how the performance of data assimilation depends on the specific set of shells being observed. In the following two subsections, we address these questions.

In our parameter setting, the turnover time of the slowest scale is $\tau_0 \approx 0.5$, while at the fastest dissipative cutoff scale, it decreases to approximately $\tau_{15} = 0.002 \tau_0 =100dt$. Sampling at these respective rates (that is, using $\Delta t_{\textit{obs}} = \tau_n$) provides discrete or (quasi-) continuous-time information on inertial range dynamics.

In the first set of experiments, observations are taken from shells $n=6,7,8$, using an observation interval of $\Delta t_{\textit{obs}} = \tau_{15}$ and an ensemble size of $L=1000$ members.
\hyperref[fig:dinamic_evol_enkf]{Figure~\ref*{fig:dinamic_evol_enkf}(a)} provides a qualitative overview of the assimilation performance, showing the temporal evolution of the EnKF-based estimates for shells $n=1$, $6$, and $13$. The ensemble estimates are continuously updated across all scales, rapidly synchronizing the directly observed $u_6$ (middle panel), whose ensemble spread quickly converges within the measurement error bounds. Dynamical correlations also enable full synchronization of distant, unobserved shells such as fast $u_{13}$ and $u_1$, albeit with a slight delay. With real-part synchronization already established (\hyperref[fig:dinamic_evol_enkf]{Fig.~\ref*{fig:dinamic_evol_enkf}(b)}), the narrow PDF of $\theta - \tilde{\theta}$ (\hyperref[fig:dinamic_evol_enkf]{Fig.~\ref*{fig:dinamic_evol_enkf}(c)}) completes the reconstruction picture by confirming phase alignment, which in turn implies synchronization of the imaginary components as well.
\begin{figure*}[]
    \centering
    \includegraphics[width=1\textwidth]{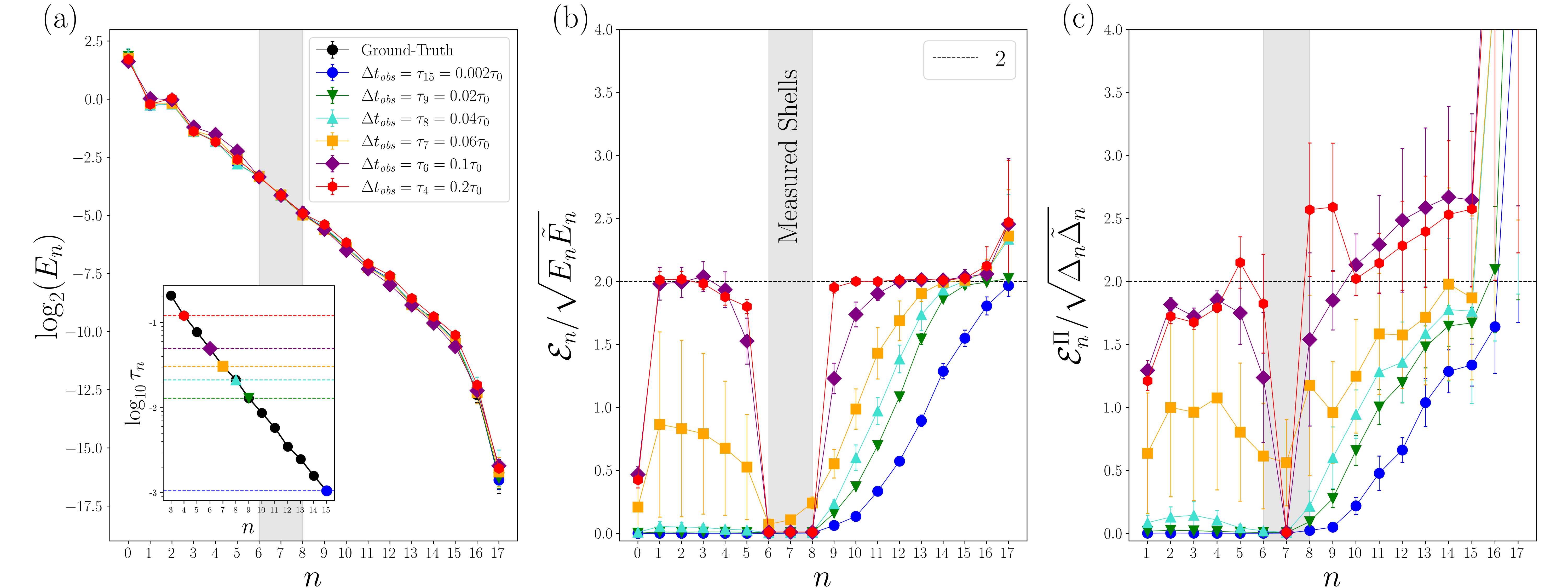}
    \caption{(a) Ground truth energy $E_n = \langle |u_n|^2 \rangle_T$ (black dots beneath the colored points) and estimated energy $\tilde{E}_n = \langle |\tilde{u}_n|^2 \rangle_{T,L}$ for different assimilation windows $\Delta t_{\textit{obs}}$. The inset shows the corrispondence with scale turnover times $\tau_n$. The grey shaded area highlights the measured scales 6, 7, and 8. (b) Synchronization error $\mathcal{E}_n$ (\hyperref[eqn:error_usual]{Eq.(\ref*{eqn:error_usual})}) normalized with energies, for various observation frequencies, with a horizontal dashed line indicating the baseline for statistical independence on non-assimilated scales. (c) Flux-based error $\mathcal{E}^\Pi_n$ (\hyperref[eqn:error_flux]{Eq.(\ref*{eqn:error_flux})}) normalized (see \hyperref[evaluation_metrics]{Evaluation Metrics} for details on normalization). All quantities displayed (including error bars) are averaged over $N_{\text{exp}} = 16$ experiments, as described in \hyperref[evaluation_metrics]{Evaluation Metrics}.}
    \label{fig:set_freq}
\end{figure*}
In \hyperref[fig:all_together]{Fig.~\ref*{fig:all_together}}, we summarize the main quantitative results of this first DA experiment. First, note that the EnKF accurately reproduces the statistical properties of the ground truth, as evidenced in \hyperref[fig:all_together]{Fig.~\ref*{fig:all_together}(a)} by the perfect overlap between the ground-truth energy spectrum $E_n$ and its EnKF-based estimate $\tilde{E}_n$.

More interestingly, the instantaneous reconstruction is also reasonably accurate (\hyperref[fig:all_together]{Fig.~\ref*{fig:all_together}(a)}). This is assessed by comparing the energy spectrum $E_n$ with the reconstruction $\ell^2$-error $\mathcal{E}_n$, as defined in \hyperref[eqn:error_usual]{Eq.~(\ref*{eqn:error_usual})}. As shown, the reconstruction is nearly perfect on the observed shells (gray region), and remains highly accurate for all slower shells ($n < 6$). For faster scales, good reconstruction is achieved up to $n = 16$, where $\mathcal{E}_n < E_n$, while at higher shell numbers ($n \ge 17$), the reconstruction becomes purely statistical with $\mathcal{E}_n \approx 2 E_n$. The quality of the instantaneous reconstruction is further quantified in \hyperref[fig:all_together]{Fig.~\ref*{fig:all_together}(b)}, where we plot the dimensionless $\ell^2$-error $\mathcal{E}_n/\sqrt{E_n \tilde{E}_n}$, normalized by the energy spectra of the ground truth and assimilated fields (\hyperref[eqn:mean_energies]{Eq.~(\ref*{eqn:mean_energies})}). Complete synchronization with the ground truth, i.e. a practically vanishing normalized error, is achieved up to the fastest observed shell ($n=8$), followed by a partially synchronized range ($9 \le n \le 16$). \hyperref[fig:all_together]{Panels~(c) and (d)} provide a qualitative overview of the time-dependent performance of data assimilation, showing both the normalized energy, $k_n^{2/3}|u_n|^2$, and the energy flux, $\Pi_n$ (see \hyperref[eqn:flux]{Eq.~(\ref*{eqn:flux})}), up to the last inertial shell ($n = 13$), over a time window encompassing several energy bursts propagating from large to small scales. The left, central, and right panels, respectively, display the ground-truth quantity, a randomly selected ensemble member, and their pointwise difference. A key observation is that, although the presence of dark halos for $1 \leq n \leq 8$ (third column of \hyperref[fig:all_together]{Panel~(c)}) confirms the accurate instantaneous tracking of both measured and large-scale dynamics, the most pronounced desynchronization events---evident in the same panel and even more so in the highly intermittent flux shown in \hyperref[fig:all_together]{Panel~(d)}---are strongly correlated with intense energy bursts, predicted with high temporal fidelity but limited magnitude accuracy.

Overall, the results are highly encouraging: the EnKF provides excellent reconstruction, even for unobserved slow scales, provided that sufficiently frequent observations are available at mesoscales. To analyze this further, \hyperref[fig:extreme_events]{Fig.~\ref*{fig:extreme_events}(a)} focuses on shell \( n = 13 \), which, while still within the inertial range, lies very close to the dissipative scales (\( n \gtrsim 15 \)). This makes it one of the most challenging shells to assimilate, due to its strongly non-Gaussian and intermittent behavior (see the lower panel of \hyperref[fig:dinamic_evol_enkf]{Fig.~\ref*{fig:dinamic_evol_enkf}}).
The scatter plot of the mean squared error $\langle |u_{13} - \tilde{u}_{13}|^2 \rangle_L$ confirms a good reconstruction in most realizations, as most points fall below the statistical baseline represented by the black line at $\langle |u_{13} - \tilde{u}_{13}|^2 \rangle_L = 2 |u_{13}|^2$. This threshold corresponds to a situation where the ensemble reproduces the correct energy $\langle|\tilde{u}_{13}|^2\rangle_L = |u_{13}|^2$, but lacks correlation with the instantaneous truth, i.e. there is no effective assimilation. A more detailed view is provided in \hyperref[fig:extreme_events]{Fig.~\ref*{fig:extreme_events}(b)}, which shows box plots of the ensemble errors $|u_{13} - \tilde{u}_{13}^{(j)}|^2$ for five different amplitude levels of $|u_{13}|$. The near overlap between the mean and median values suggests that the ensemble error distribution is nearly symmetric and is not significantly affected by outliers.
These findings are consistent with the lower panel of \hyperref[fig:dinamic_evol_enkf]{Fig.~\ref*{fig:dinamic_evol_enkf}(a)}, where the ensemble is seen to symmetrically span the true value. For example, for states with $\Re(u_{13}) > 0$, the ensemble members include values with $\Re(\tilde{u}_{13}^{(j)}) < 0$. This behavior confirms that an ensemble size of $L=1000$ is sufficient to sample the relevant statistical space of the system.

In the following two subsections, we systematically investigate the role of $\Delta t_{\textit{obs}}$ and the choice of measured shells on both the statistical and instantaneous synchronization processes.
%%%%%%%%%%%%%%%%%%%%%%%%%%%%%%

%%%%%%%%%%%%% Subsection EnKF sparseness %%%%%%%%%%%%%
%%%%%%%%%%%%%%%%%%%%%%%%%%%%%%%%%%%%%%%%%%%%%%%%%%%%
\vspace{-0.5cm}
\subsection{Performance Against Measurements Frequency}
\label{sub:sparseness}
\hyperref[fig:set_freq]{Figure~\ref*{fig:set_freq}(a)} shows that varying the observation interval $\Delta t_{\textit{obs}}$ within the range $\tau_{15} \le \Delta t_{\textit{obs}} \le \tau_{4}$ does not substantially affect the ability of the EnKF to recover the statistical shape of the ground-truth energy spectrum $E_n = \langle |u_n|^2 \rangle_T$ through its estimate $\tilde{E}_n = \langle |\tilde{u}_n|^2 \rangle_{T,L}$. The inset indicates how the selected values of $\Delta t_{\textit{obs}}$ relate to the turnover times of the various shells. However, differences arise when examining the instantaneous reconstruction through the normalized error $\mathcal{E}_n / \sqrt{E_n \tilde{E}_n}$, as shown in \hyperref[fig:set_freq]{Figure~\ref*{fig:set_freq}(b)}. Discrepancies begin to appear on unmeasured shells as soon as the observation interval exceeds the turnover time of shell $n=9$, i.e., when $\Delta t_{\textit{obs}} > \tau_9$. This metric, introduced in \hyperref[sec:experimental_setup]{Section~\ref*{sec:experimental_setup}}, is compared with a baseline value of 2 (dashed horizontal line), which denotes the threshold above which assimilation no longer produces an effective correction toward the ground truth. Only scales with normalized error below 2 can be successfully assimilated.

\hyperref[fig:set_freq]{Figure~\ref*{fig:set_freq}(c)} further supports these findings by displaying the instantaneous reconstruction of the energy flux across scales. It shows that reducing the measurement frequency (that is, increasing $\Delta t_{\textit{obs}}$) perturbs the cascade and disrupts flux synchronization, as captured by the flux error metric $\mathcal{E}^{\Pi}_n / \sqrt{\Delta_n \tilde{\Delta}_n}$ (see \hyperref[evaluation_metrics]{Evaluation Metrics}). In this regime, the assimilation updates act effectively as an external forcing, disturbing both the measured shells and their neighbors, and leading to the formation of an energy bottleneck within the cascade (see also the discussion in the next \hyperref[sub:scale_schemes]{subsection}).

We note that since the ensemble is updated only at discrete measurement times, the number of posterior estimates is directly proportional to the length of the assimilation window $\Delta t_{\textit{obs}}$; shorter windows naturally provide more data points for averaging. After the start of the assimilation experiment, a transient period of $T = 10\tau_0$ is used to allow the ensemble to reach a statistically steady state (and, when applicable, to synchronize with the ground truth) before the averaging begins.  
As a result, all computed averages reflect stationary conditions for both the ensemble and the ground truth, and the associated error bars inherently account for uncertainties related to the varying number of available samples.\\
The magnitude of the error bars also reflects the sensitivity to the inflation factor (as discussed in \hyperref[sub:inflation]{Subsection~\ref*{sub:inflation}}) used for stabilization. For example, the curve corresponding to $\tau_{7}$ in \hyperref[fig:set_freq]{Fig.~\ref*{fig:set_freq}(b)} exhibits a pronounced increase in error-bar length, estimated as in \hyperref[eqn:error_bars]{Eq.~(\ref*{eqn:error_bars}}), at scales beyond those directly observed, signaling that assimilation was less stable in those cases. This pattern aligns with the theoretical insights of Gottwald and Majda~\cite{gottwald_filter_divergence}, who showed that the stability of the filter critically depends on the observation interval $\Delta t_{\textit{obs}}$. Specifically, when observations are too frequent relative to the relaxation rate toward the attractor, the forecast step is too short for the model dynamics to evolve to overly constrained and potentially unrealistic states. Conversely, when observations are too sparse, the ensemble can spread sufficiently to capture the attractor structure. In other words, while filter divergence is unlikely at either extreme (very small or very large $\Delta t_{\textit{obs}}$) there exists an intermediate regime where the ensemble may be neither sufficiently constrained by observations nor sufficiently dispersed to maintain dynamical consistency, resulting in degraded performance and potential divergence.

The total errors are summarized in \hyperref[tab:measurement_sparseness]{Table~\ref*{tab:measurement_sparseness}} by summing the errors from the first scale (excluding the forced one) up to the dissipative cutoff at scale 15:
$\mathcal{E} = \sum_{n=1}^{15} \mathcal{E}_n / \sqrt{\tilde{E}_n E_n}$ and $\mathcal{E}^\Pi = \sum_{n=1}^{15} \mathcal{E}_n^\Pi / \sqrt{\tilde{\Delta}_n \Delta_n}$. A similar procedure is applied to the error bars to obtain the total uncertainties $\Delta\mathcal{E}$ and $\Delta\mathcal{E}^\Pi$.
Furthermore, the last column of \hyperref[tab:measurement_sparseness]{Table~\ref*{tab:measurement_sparseness}} reports the number of independent assimilation experiments (out of $N{_\text{exp}} = 16$) that required the correct adjustment of the inflation strength factor $\lambda$, as defined in \hyperref[eqn:inflation_coeff]{Eq.~(\ref*{eqn:inflation_coeff})}.

The combined analysis of \hyperref[tab:measurement_sparseness]{Table~\ref*{tab:measurement_sparseness}} and \hyperref[fig:set_freq]{Fig.~\ref*{fig:set_freq}(b)} leads to the conclusion that there exists a measurement frequency threshold that enables synchronization of all scales slower than those directly observed, provided that the measurement frequency exceeds the turnover time of the fastest measured shell. Specifically, measurements at $\tau_9$ yield complete synchronization of all larger scales.
\begin{table}[h!]
\begin{ruledtabular}
    \begin{tabular}{cccc}
    6, 7, 8& $\mathcal{E}\pm\Delta\mathcal{E}$ & $\mathcal{E}_{\Pi}\pm\Delta\mathcal{E}_{\Pi}$ &
    $N_{\text{inflation}}$\\
    \hline
$\Delta t_{\text{obs}}=\tau_{15} = 0.002\tau_0$ & $4.85 \pm 0.26$  & $5.11 \pm 0.85$  & $1$ \\
$\Delta t_{\text{obs}}=\tau_{9} = 0.02\tau_0$   & $7.75 \pm 0.17$  & $8.13 \pm 0.98$  & $1$ \\
$\Delta t_{\text{obs}}=\tau_{8} = 0.04\tau_0$   & $9.10 \pm 0.69$  & $10.07 \pm 3.21$ & $6$ \\
$\Delta t_{\text{obs}}=\tau_{7} = 0.06\tau_0$   & $14.68 \pm 3.90$ & $17.76 \pm 8.35$ & $8$ \\
$\Delta t_{\text{obs}}=\tau_{6} = 0.1\tau_0$    & $22.44 \pm 1.14$ & $27.89 \pm 5.20$ & $0$ \\
$\Delta t_{\text{obs}}=\tau_{4} = 0.2\tau_0$    & $23.73 \pm 0.50$ & $29.49 \pm 4.23$ & $0$ \\
    \end{tabular}
    \end{ruledtabular}
    \label{tab:measurement_sparseness}
    \caption{The total error is computed by summing the normalized errors from the first scale $n=1$ (excluding the forced one) up to the dissipative cutoff at scale $n=15$: $\mathcal{E} = \sum_{n=1}^{15} \mathcal{E}_n / \sqrt{\tilde{E}_n E_n}$ and $\mathcal{E}_\Pi = \sum_{n=1}^{15} \mathcal{E}_n^\Pi / \sqrt{\tilde{\Delta}_n \Delta_n}$. A similar summation is applied to the error bars to obtain $\Delta\mathcal{E}$ and $\Delta\mathcal{E}^\Pi$.
Details on the number of experiments (out of $N_{\text{exp}} = 16$) that required proper tuning of the inflation strength factor are also provided.}
\end{table}
%%%%%%%%%%%%% End Subsection EnKF sparseness %%%%%%%%%%%%%
%%%%%%%%%%%%%%%%%%%%%%%%%%%%%%%%%%%%%%%%%%%%%%%%%%%%

%%%%%%%%%%%%% Subsection EnKF Measured Scales %%%%%%
%%%%%%%%%%%%%%%%%%%%%%%%%%%%%%%%%%%%%%%%%%%%%%%%%%%%
\vspace{-0.5cm}
\subsection{Measured Scale Schemes}
\label{sub:scale_schemes}
In this work we perform a systematic investigation of how the choice of measured scales affects both statistical and instantaneous DA reconstruction, as well as cascading energy transfer. After establishing a synchronization threshold based on observation sparseness $\Delta t_{\textit{obs}}$, we fix it to $\tau_{15}$ (the turnover time of the dissipative scale), which so far has ensured the best assimilation performance, and vary the set of measured shells to assess its impact.
\begin{figure*}[]
    \centering
    \includegraphics[width=1\textwidth]{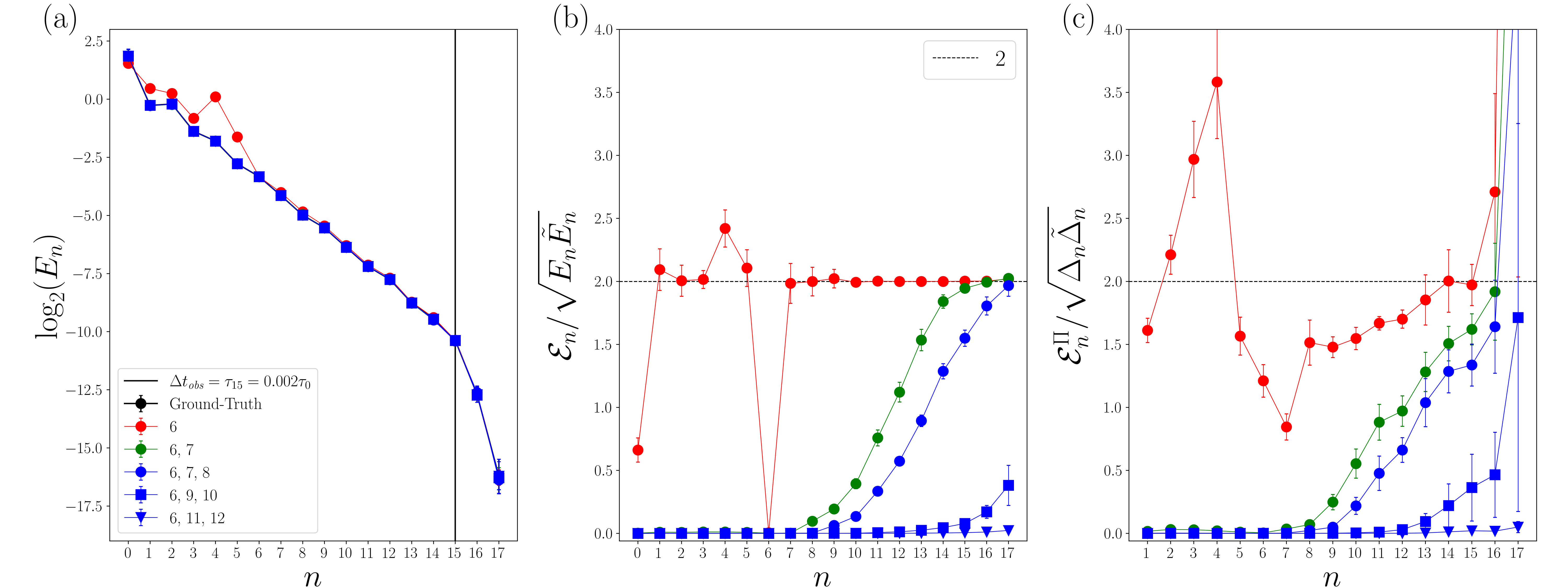}
    \caption{(a) Ground truth energy $E_n=\langle |u_n|^2 \rangle_T$ (black dots beneath the colored points) and estimated energy $\tilde{E}_n=\langle |\tilde{u}_n|^2 \rangle_{T,L}$ for different sets of measured shells; the vertical solid line indicates the shell turnover time associated with the assimilation window $\Delta t_{\textit{obs}}$. (b) Synchronization error $\mathcal{E}_n$ (\hyperref[eqn:error_usual]{Eq.(\ref*{eqn:error_usual})}) normalized with energies, with a horizontal dashed line indicating the baseline for statistical independence on non-assimilated scales. (c) Flux-based error $\mathcal{E}^\Pi_n$ (\hyperref[eqn:error_flux]{Eq.(\ref*{eqn:error_flux})}) normalized (see \hyperref[evaluation_metrics]{Evaluation Metrics} for details on normalization). All quantities displayed (including error bars) are averaged over $N_{\text{exp}} = 16$ experiments, as described in \hyperref[evaluation_metrics]{Evaluation Metrics}.}
    \label{fig:set_shells}
\end{figure*}
A central finding is that measuring two adjacent shells within the inertial range, such as shells 6 and 7, suffices to achieve full synchronization of the large-scale dynamics (\hyperref[fig:set_shells]{Fig.~\ref*{fig:set_shells}(b)}). Adding a third measurement (e.g., shells 6, 7, and 8) only assimilates an additional fast scale without significantly enhancing synchronization. In contrast, measuring a single shell (e.g., shell $6$) fails to synchronize the inertial range and disrupts the energy cascade. This disruption arises from the structure of the energy flux, which is mediated through triadic couplings involving terms such as $u_n^* u_{n+1}^* u_{n+2}$ and $u_{n-1}^* u_n^* u_{n+1}$. An inadequate reconstruction of these nonlinear interactions leads to bottlenecks, blocking the transfer of energy from the forced scales and causing its accumulation in the shells immediately preceding the measured one (see \hyperref[fig:set_shells]{Fig.~\ref*{fig:set_shells}(a)}).
Contiguity between measured shells is therefore critical: adjacent measurements fully constrain the local flux, ensuring seamless energy transfer across scales. 

Measurement of three non-consecutive shells, such as 6–9–10 or 6–11–12 (\hyperref[fig:set_shells]{Fig.~\ref*{fig:set_shells}(b) and (c)}), substantially improves the assimilation of faster scales (see \hyperref[tab:measured_scale_schemes]{Table~\ref*{tab:measured_scale_schemes}}), even beyond the dissipative range, without adverse effects due to gaps. However, excluding a large- or mesoscale shell severely destabilizes the EnKF scheme: configurations measuring only fast scales (e.g., shells 11–12 or 13–14) exhibit blow-up in over $90\%$ of cases. This instability cannot be suppressed by inflation alone and would require prohibitive ensemble sizes or more sophisticated techniques, which are beyond the scope of this work.

The origin of this instability can be clarified by considering the simpler case where only a single shell is measured, case where the EnKF update reduces to
\begin{equation}
    \label{eqn:kf_scheme}
    \tilde{U}_n^{(j)} = \hat{U}_n^{(j)} + \frac{\sigma_{nm}}{\sigma_{mm}^2 + \sigma^2} \left[Z_{m} - \hat{U}_{m}^{(j)} - V^{(j)}\right]\ ,
\end{equation}

where $\sigma^2$ is the variance of the observation error, $\sigma_{mm}^2$ the variance of the measured shell, and $\sigma_{mn}$ the cross-correlation with the shell $n$. When fast scales are measured, $\sigma^2$ grows with $\langle |u_{m}|^2 \rangle_T$, and burst events amplify the innovation term, leading to excessively large corrections that inject energy forward and backward throughout the cascade, ultimately destabilizing the ensemble.

This instability can be mitigated by also measuring a large- or mesoscale shell: for instance, adding shell 6 reduces the blow-up rate from $\sim90\%$ to $\sim13\%$, with further stabilization achieved by appropriately tuning the inflation factor $\lambda$ (see \hyperref[sub:inflation]{Subsection~\ref*{sub:inflation}} for explanations). Measuring a large- or mesoscale shell promotes the synchronization of traveling bursts originating from large scales and enhances the overall stability of the DA process, consistently with theoretical results for dissipative systems, where a finite number of determining modes suffices to control long-term dynamics~\cite{olson_titi_determining_modes,azouani_titi}, and, for example, in the Navier–Stokes context, measuring continuously up to $k=0.2k_{\eta}$ (where $k_{\eta}$ is the dissipative wavenumber) enables the recovery of small-scale dynamics independently of the DA method used~\cite{yoshida_data_assimilation,li_small_scale_reconstruction,patricio}.

All the results presented in this section are summarized in \hyperref[tab:measured_scale_schemes]{Table~\ref*{tab:measured_scale_schemes}} by summing the errors from the first scale (excluding the forced one) up to the dissipative cutoff at scale 15:
$\mathcal{E} = \sum_{n=1}^{15} \mathcal{E}_n / \sqrt{\tilde{E}_n E_n}$, and $\mathcal{E}_\Pi = \sum_{n=1}^{15} \mathcal{E}_n^\Pi / \sqrt{\tilde{\Delta}_n \Delta_n}$.  
A similar procedure is applied to the error bars to obtain the total uncertainties $\Delta\mathcal{E}$ and $\Delta\mathcal{E}^\Pi$, consistently following standard rules for the propagation of experimental errors.  
In addition, detailed information is provided on the number of experiments (out of $N_{\text{exp}} = 16$) that required the appropriate adjustment of the inflation strength factor $\lambda$ (see \hyperref[sub:inflation]{Subsection~\ref*{sub:inflation}}).
\begin{table}[h!]
    \begin{ruledtabular}
    \begin{tabular}{cccc}
    $\Delta t_{\text{obs}}=\tau_{15}=0.002\tau_0 $ & $\mathcal{E}\pm\Delta\mathcal{E}$ & $\mathcal{E}_{\Pi}\pm\Delta\mathcal{E}_{\Pi}$ &
    $N_{\text{inflation}}$\\
    \hline
%6, 14         & $26.83 \pm 1.21$ & $28.40 \pm 3.23$ & $0$ \\
%4, 14         & $25.81 \pm 1.88$ & $28.74 \pm 4.26$ & $0$ \\
%2, 14         & $23.15 \pm 1.36$ & $27.71 \pm 4.01$ & $0$ \\
6             & $28.65 \pm 1.05$ & $27.74 \pm 2.47$ & $0$ \\
6, 7          & $7.94 \pm 0.43$  & $7.29 \pm 0.95$  & $8$ \\
6, 7, 8       & $4.85 \pm 0.26$  & $5.11 \pm 0.85$  & $1$ \\
6, 9, 10      & $0.18 \pm 0.04$  & $0.73 \pm 0.53$  & $2$ \\
6, 11, 12     & $0.016 \pm 0.004$ & $0.04 \pm 0.03$ & $3$ \\
    %11, 12 & 0.015\pm0.000 & 0.029\pm0.00 & 14 (87.5\%) *\\
    %12, 13 & 0.009\pm0.000 & 0.016\pm0.00 & 14 (87.5\%) *\\
    %13, 14 & 24.67\pm0.000 & 71.25\pm0.00 & 16 (100\%) *\\
    \end{tabular}
    \end{ruledtabular}
    \caption{The total error is computed by summing the normalized errors from the first scale $n=1$ (excluding the forced one) up to the dissipative cutoff at scale $n=15$: $\mathcal{E} = \sum_{n=1}^{15} \mathcal{E}_n / \sqrt{\tilde{E}_n E_n}$ and $\mathcal{E}_\Pi = \sum_{n=1}^{15} \mathcal{E}_n^\Pi / \sqrt{\tilde{\Delta}_n \Delta_n}$. A similar summation is applied to the error bars to obtain $\Delta\mathcal{E}$ and $\Delta\mathcal{E}^\Pi$.
Details on the number of experiments (out of $N_{\text{exp}} = 16$) that required proper tuning of the inflation strength factor are also provided.}
    \label{tab:measured_scale_schemes}
\end{table}
%%%%%%%%%%%%% End Subsection EnKF Measured Scales %%%%%%
%%%%%%%%%%%%%%%%%%%%%%%%%%%%%%%%%%%%%%%%%%%%%%%%%%%%

%%%%%%%%%%%%% Section Comparisongs with Nuding %%%%%%%%%%%%%
%%%%%%%%%%%%%%%%%%%%%%%%%%%%%%%%%%%%%%%%%%%%%%%%%%%%%%%%%%%%%%%
\begin{figure*}[]
    \centering
    \begin{minipage}[t]{0.544\textwidth}
        \centering
        \includegraphics[width=\textwidth]{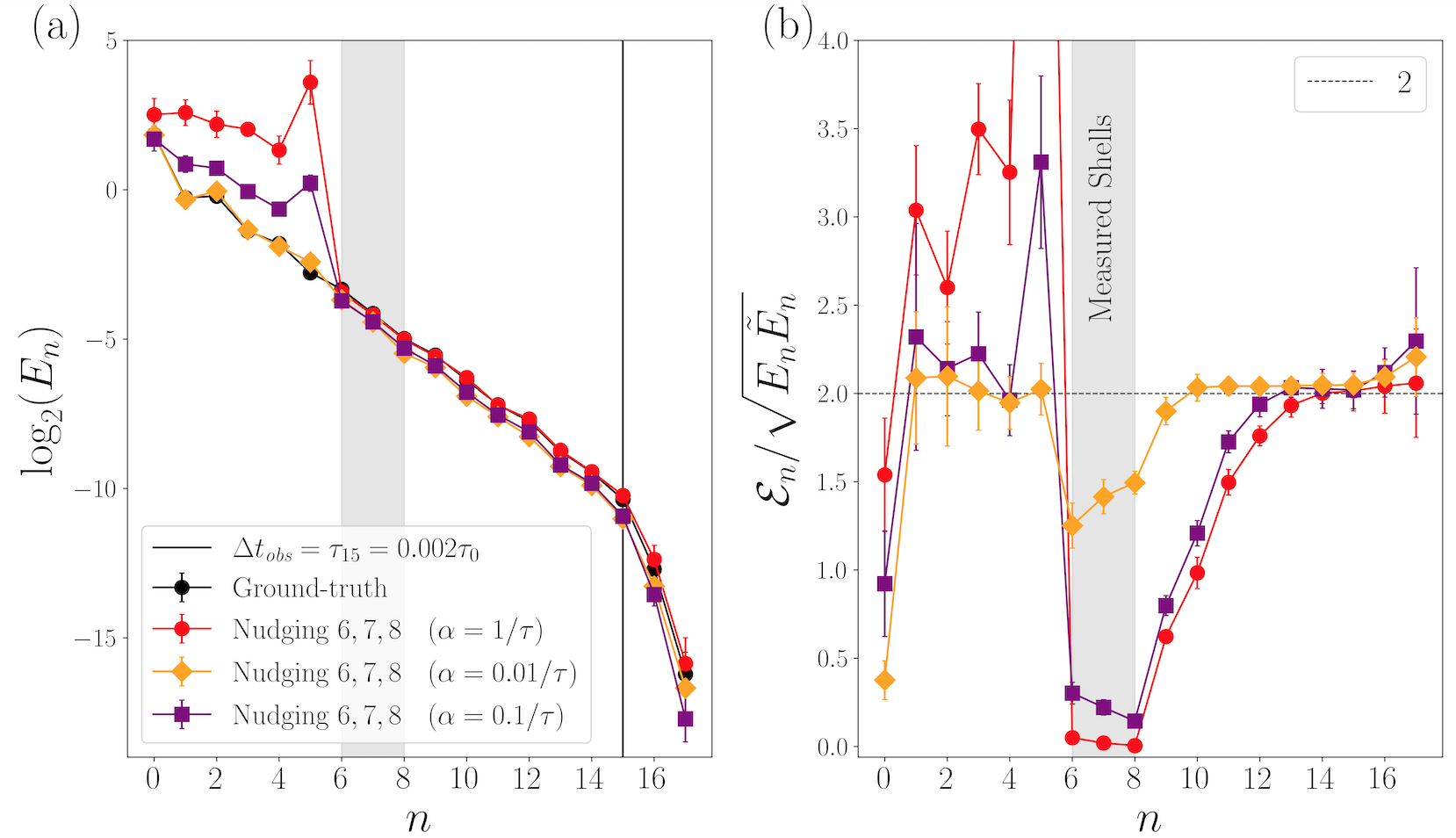}
        \caption{(a) Ground truth energy $E_n=\langle |u_n|^2 \rangle_T$ and estimated energy with Nudging $\tilde{E}_n=\langle |\tilde{u}_n|^2 \rangle_{T}$, at varying nudging coefficient $\alpha=\alpha'/\tau$, with $\tau=\Delta t_{\textit{obs}}=\tau_{15}$; the vertical solid line indicates the shell turnover time associated with the assimilation window $\Delta t_{\textit{obs}}$. (b) Synchronization error $\mathcal{E}_n$ (\hyperref[eqn:error_usual]{Eq.(\ref*{eqn:error_usual})}) normalized with energies, with a horizontal dashed line indicating the baseline for statistical independence on non-assimilated scales. All quantities displayed (including error bars) are averaged over $N_{\text{exp}} = 16$ experiments, as described in \hyperref[evaluation_metrics]{Evaluation Metrics}.}
        \label{fig:nudging_alpha_partial}
    \end{minipage}
    \hfill
    \begin{minipage}[t]{0.4465\textwidth}
        \centering
        \includegraphics[width=\textwidth]{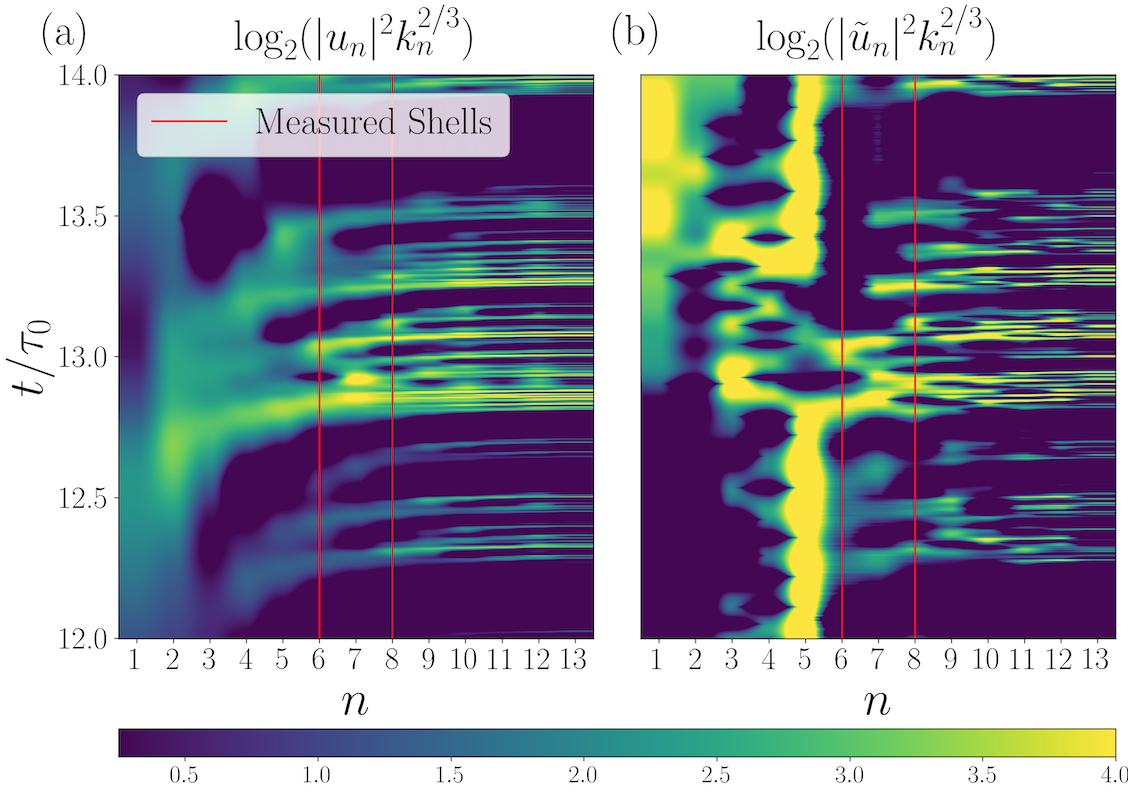}
        \caption{(a) Normalized ground-truth energy $|u_n|^2 k_n^{2/3}$, including signal bursts during the interval $12 \leq t / \tau_0 \leq 14$. Red vertical lines indicate the measured shells ($n = 6$, $7$, and $8$). (b) Normalized energy of the Nudging estimate $|\tilde{u}_n|^2 k_n^{2/3}$.}
        \label{fig:energy_evolution_nudging}
    \end{minipage}
\end{figure*}
\section{Comparisons with Nudging}
\label{sec:comparisons_nudging}
Nudging acts as a Newtonian relaxation process, wherein the system state estimate $\tilde{\bm{u}}$ is gradually pulled toward the measurements $\bm{z}$ by introducing a feedback term into the governing equations:
\begin{equation} 
    \label{eqn:nudging}
    \left(\frac{d}{dt} + \nu k_n^2\right)\tilde{u}_n(t) = G_n[\bm{u}] + f_n + \alpha_n \delta_{nm} \left[T_n(\bm{z},t) - \tilde{u}_n(t)\right]\,, 
\end{equation} 
where $\alpha_n$ controls the feedback strength, and $\delta_{nm}$ ensures corrections are applied only to the observed variables $m$ (details on the integration scheme used for \hyperref[eqn:nudging]{Eq.(\ref*{eqn:nudging})} are provided in \hyperref[app:rk4]{Appendix~\ref*{app:rk4}}). Larger $\alpha_n$ values accelerate convergence but can amplify noise, whereas smaller values produce smoother corrections, which is particularly advantageous in noisy scenarios.

Given that measurements are not available at every time step, a continuous linear interpolation $T_n(\bm{z},t)$ is introduced:
\begin{equation}
T_n(\bm{z}, t) = \frac{(t_{k+1} - t) z_n(t_k) + (t - t_k) z_n(t_{k+1})}{t_{k+1} - t_k},
\end{equation}
where $z_n(t_k) = u_n(t_k) + \epsilon_n(t_k)$ and $z_n(t_{k+1}) = u_n(t_{k+1}) + \epsilon_n(t_{k+1})$ are the ground-truth measurements at times $t_k$ and $t_{k+1}$, respectively. The terms $\epsilon_n(t_k)$ and $\epsilon_n(t_{k+1})$ are independent measurement errors drawn from a complex zero-mean Gaussian distribution, with covariance $(R_\mathbb{C})_{mn} = \delta_{mn}\ 2\ (0.05)^2 \langle |u_n|^2 \rangle$, where $m,n = 0,\ldots,N-1$ index the measured shells (this corresponds to the complex version of the matrix in \hyperref[eqn:meas_cov_matrix]{Eq.~(\ref*{eqn:meas_cov_matrix}})).

Nudging applies localized corrections by adjusting only the measured variables, without propagating information to unobserved components. This is equivalent to assuming a fully diagonal implicit correlation structure. Furthermore, since nudging lacks a mechanism to estimate the forecast uncertainty over time, the coupling parameter $\alpha_n$ remains fixed, limiting the adaptability in dynamically evolving systems. Although several methods have been proposed to estimate $\alpha_n$~\cite{zou_optimal_nudging_1992,stauffer_optimal_nudging_1993,pazo_data_assimilation_2016}, in our formulation this coefficient remains the only free parameter and must therefore be tuned empirically.

To ensure that the relaxation rate towards the ground truth scales with observation sparseness, we define the coupling coefficient as $\alpha_n = \alpha'_n / \tau$, where $\tau = \Delta t_{\textit{obs}}$. In all experiments using $\Delta t_{\textit{obs}} = \tau_{15}$, the assimilation performance was found to depend only weakly on the specific value of $\alpha'_n$ on each scale. Therefore, we set $\alpha'_n = \alpha'$ uniformly across all scales. Once this is decided, the way we optimize $\alpha'$ depends on the specific set of measured shells.
\begin{figure*}[t]
    \centering
    \includegraphics[width=1\textwidth]{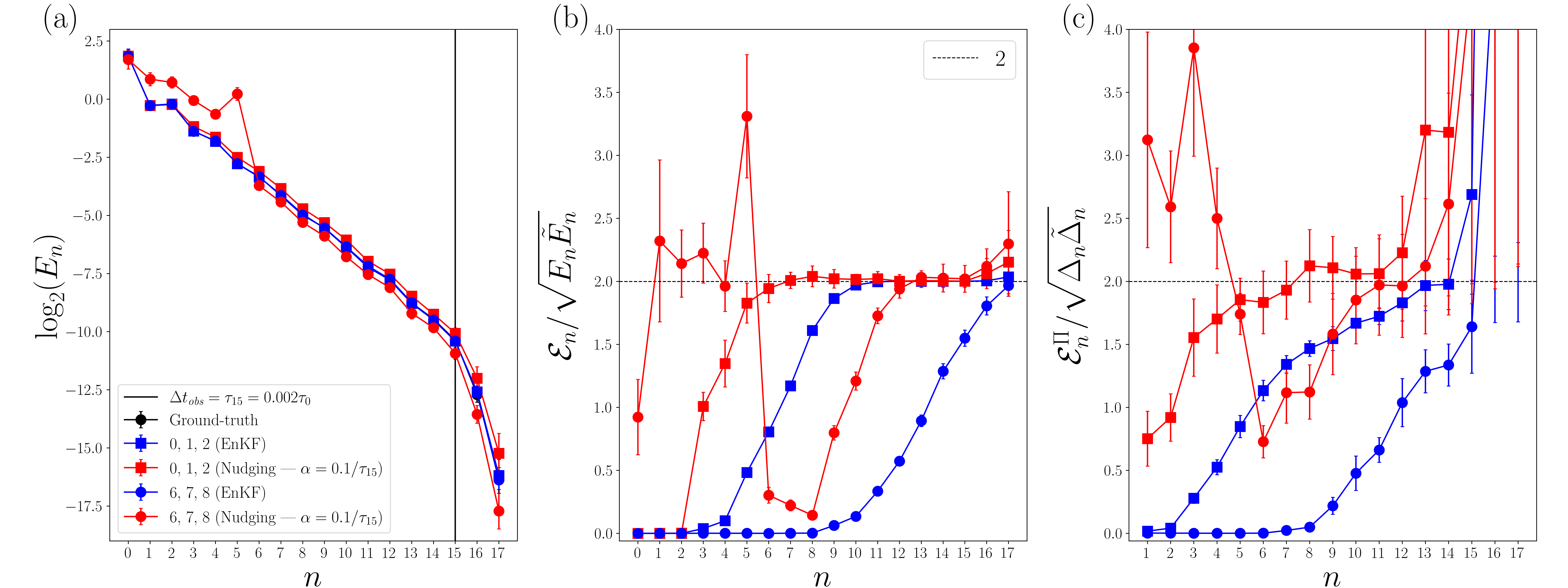}
    \caption{Comparison between Nudging and EnKF. (a) The structure functions for the labelled cases, for ground-truth $\langle |u_n|^2 \rangle_T$ (black dots beneath the colored points), nudged solution $\langle |\tilde{u}_n|^2\rangle_{T}$ and EnKF $\langle |\tilde{u}_n|^2\rangle_{T,L}$; the vertical solid line indicates the shell turnover time associated with the assimilation window $\Delta t_{\textit{obs}}=\tau_{15}=0.002\tau_0$. (b) Synchronization error $\mathcal{E}_n$ (\hyperref[eqn:error_usual]{Eq.(\ref*{eqn:error_usual})}) normalized with energies, with a horizontal dashed line indicating the baseline for statistical independence on non-assimilated scales. (c) Flux-based error $\mathcal{E}^\Pi_n$ (\hyperref[eqn:error_flux]{Eq.(\ref*{eqn:error_flux})}) normalized (see \hyperref[evaluation_metrics]{Evaluation Metrics} for details on normalization). All quantities displayed (including error bars) are averaged over $N_{\text{exp}} = 16$ experiments, as described in \hyperref[evaluation_metrics]{Evaluation Metrics}.}
    \label{fig:comparison_nudging}
\end{figure*}
\hyperref[fig:nudging_alpha_partial]{Figure~\ref*{fig:nudging_alpha_partial}(a)} illustrates the effect of varying $\alpha'$ on the estimate of the nudging-based energy spectrum estimate $\tilde{E}_n = \langle |\tilde{u}_n|^2 \rangle_T$, compared with the ground truth spectrum $E_n = \langle |u_n|^2 \rangle_T$, in the case where the shells $n=6,7,8$ are measured. It can be seen that the extreme locality of nudging corrections leads to energy accumulation just before the measured scales, with the magnitude of the effect increasing with $\alpha'$. This behavior also reflects in the degradation of the instantaneous reconstruction shown in \hyperref[fig:nudging_alpha_partial]{Fig.~\ref*{fig:nudging_alpha_partial}(b)}. Combining these two results, we observe that for small $\alpha'$ values the energy cascade remains undisturbed, but no effective assimilation occurs. In contrast, high values of $\alpha'$ improve reconstruction but lead to accumulation of unphysical energy in the spectrum (see \hyperref[fig:energy_evolution_nudging]{Fig.~\ref*{fig:energy_evolution_nudging}(b)}). As a compromise, we identify $\alpha' = 0.1$ as the "optimal" value.

We recall that in the Nudging procedure, we do not use an ensemble but only a single forecast. Therefore, initialization is performed using \hyperref[eqn:initial_guess_shell_condition_enkf]{Eq.~(\ref*{eqn:initial_guess_shell_condition_enkf})}, which reduces to the same expression without dependence on the ensemble index $j$.

\hyperref[fig:comparison_nudging]{Figure~\ref*{fig:comparison_nudging}} presents a final comparison between EnKF and Nudging for the cases where measurements are taken on the sets $n = 6,7,8$ and $n = 0,1,2$, using $\Delta t_{\textit{obs}} = \tau_{15}$. The statistical reconstruction of the ground truth, shown in \hyperref[fig:comparison_nudging]{Fig.~\ref*{fig:comparison_nudging}(a)}, indicates that while the EnKF accurately reproduces the spectral shape in both cases, Nudging leads to energy accumulation when the measured shells are $n = 6,7,8$. This is a direct consequence of the localized coupling, as previously discussed. This accumulation negatively affects the instantaneous reconstruction of the ground truth, as seen in the normalized $\ell^2$-error (defined in \hyperref[eqn:error_usual]{Eq.(\ref*{eqn:error_usual})}) and shown in \hyperref[fig:comparison_nudging]{Panel~(b)}. Strong desynchronization is also evident in the flux, reflected in the normalized flux-based error shown in \hyperref[fig:comparison_nudging]{Panel~(c)}. These results highlight the inability of Nudging—unlike that of EnKF—to infer unobserved scales from inertial range measurements. 

We also report results for the experiment where scales $n = 0,1,2$ are measured. Consistent with previous discussion, when large scales (including forced shell $n=0$) are observed continuously, energy accumulation is avoided because of immediate synchronization of large-scale bursts. In this case, the optimal choice of $\alpha'$ follows the criterion of minimizing the total MSE on the measured shells, $\sum_{n=0}^2 \langle |u_n - \tilde{u}_n|^2 \rangle_T$, again yielding $\alpha' = 0.1$. Finally, although no significant cascade perturbations are observed when $n = 0,1,2$ is measured, \hyperref[fig:comparison_nudging]{Figure~\ref*{fig:comparison_nudging}} confirms that the localized nature of nudging corrections still limits their effectiveness in reconstructing the cascade deeper in the inertial range. In contrast, the EnKF successfully infers up to five additional shells beyond those observed, reaching the independence threshold at $n = 10$, compared to $n = 5$ for Nudging. In terms of scale separation, this corresponds to a factor of $2^5 = 32$.
%%%%%%%%%%%%% End Section Comparison with Nuding %%%%%%%%%%%%%
%%%%%%%%%%%%%%%%%%%%%%%%%%%%%%%%%%%%%%%%%%%%%%%%%%%%%%%%%%%%%%%

%%%%%%%%%%%%% Section Comparison with 4D-Var %%%%%%%%%%%%%
%%%%%%%%%%%%%%%%%%%%%%%%%%%%%%%%%%%%%%%%%%%%%%%%%%%%%%%%%%%%%%
\section{Comparison with En4D-Var}
\label{sec:comparisons_4dvar}
In 4D-Var, assimilation is carried out over a finite time window, within which the observations are distributed, and the forecast trajectory is iteratively adjusted to them by minimizing a cost function. Unlike EnKF, where corrections are applied only at the observation times, the 4D-Var is a smoother that adjusts the assimilating model to observations available over an entire assimilation window, making the posterior estimate $\tilde{\bm{u}}(t)$ dependent not only on the observations but also on the window length $T_a$.

The posterior estimate $\tilde{\bm{u}}(t)$ in 4D-Var is obtained such that it minimize a cost function defined as:
\begin{eqnarray}
J &=& \left\| \hat{\bm{u}}(0) - \tilde{\bm{u}}(0) \right\|_{\mathcal{B}}^2 
\nonumber \\
&+& \sum_{t=0}^{T_a} \left\| \bm{z}(t) - \bm{H}_{\mathbb{C}} \tilde{\bm{u}}(t) - \bm{v}(t) \right\|_{\bm{R}_\mathbb{C}}^2\,,\label{Jij}
\end{eqnarray}
where \( \| \bm{a} \|_{\mathcal{L}}^2 = \bm{a}^\dagger \mathcal{L}^{-1} \bm{a} \) denotes the squared weighted \( \ell^2 \)-norm.

The variable $\hat{\bm{u}}(0)$ represents the prior (or background) state at the beginning of the window, and $\tilde{\bm{u}}(0)$ is the posterior optimized state (also known as the smoother). In the strong-constraint (i.e. perfect model framework) 4D-Var approach used here, the posterior trajectory is fully determined by its initial condition, i.e., \( \tilde{\bm{u}}(t) = \Phi_{0 \to t}(\tilde{\bm{u}}(0)) \), where \( \Phi_{0 \to t} \) is the time-marching operator associated with the nonlinear shell model governing equation.
The matrix $\mathcal{B}$ is the complex-valued background error covariance, assumed to be given, and $\bm{H}_{\mathbb{C}}$ is the complex extension of the observation operator $\bm{H}$ (see \hyperref[sec:enkf]{Sec.~\ref*{sec:enkf}}), needed to produce the complex measurements $z\in\mathbb{C}^{M}$. Finally, $\bm{v}(t)$ is a realization of complex zero-mean Gaussian noise with covariance $\bm{R}_\mathbb{C}$ (complex version of the matrix in \hyperref[eqn:meas_cov_matrix]{Eq.~(\ref*{eqn:meas_cov_matrix}})).

The cost function comprises two terms: the first penalizes the deviation of the posterior from the prior at the initial time, while the second accumulates the innovations, i.e., the discrepancies between the posterior trajectory and the observations over the assimilation window. 
Here, we adopt an incremental approach to 4D-Var in which we simplify the cost function by linearizing \( \Phi_{0 \to t} \) around an initial guess for $\tilde{\bm{u}}(0)$ \cite{courtier_4dvar_1994}. This results in a quadratic approximation of the cost function, which is then minimized using Newton's method. For implementation, we save the linearised operators \( \Phi_{0 \to t} \) during the forward time-marching, then use them to form the adjoint model, required for obtaining the Jacobian of the cost-function, and explicitly calculate the Hessian, required by the Newton's method. Once a solution to the simplified cost function is found, the process is repeated by re-linearizing \( \Phi_{0 \to t} \) around the updated solution. This iterative process continues until convergence is achieved.
\begin{figure*}[t]
    \centering
    \includegraphics[width=1\textwidth]{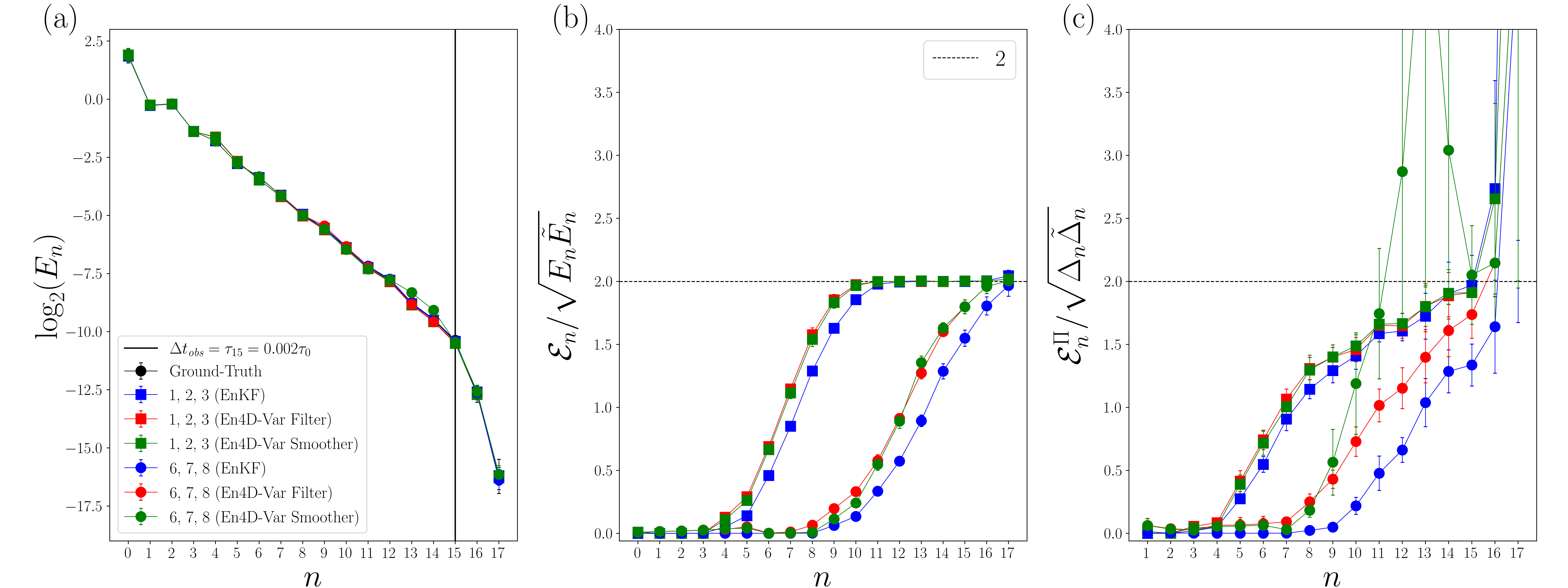} 
    \caption{Comparison between En4D-Var (both smoother and filter) and EnKF. (a) Ground truth energy $E_n=\langle |u_n|^2 \rangle_T$ (black dots beneath the colored points) and estimated energy with En4D-Var and EnKF $\tilde{E}_n=\langle |\tilde{u}_n|^2 \rangle_{T,L}$; the vertical solid line indicates the shell turnover time associated with the assimilation window $\Delta t_{\textit{obs}}$. (b) Synchronization error $\mathcal{E}_n$ (\hyperref[eqn:error_usual]{Eq.(\ref*{eqn:error_usual})}) normalized with energies, with a horizontal dashed line indicating the baseline for statistical independence on non-assimilated scales. (c) Flux-based error $\mathcal{E}^\Pi_n$ (\hyperref[eqn:error_flux]{Eq.(\ref*{eqn:error_flux})}) normalized (see \hyperref[evaluation_metrics]{Evaluation Metrics} for details on normalization). All quantities displayed (including error bars) are averaged over $N_{\text{exp}} = 16$ experiments, as described in \hyperref[evaluation_metrics]{Evaluation Metrics}.}
    \label{fig:4dvar_enfk}
\end{figure*}
Due to the chaotic nature of the shell model, any error - whether from the initial guess or from linearization - grows exponentially over time. As a result, the length of the assimilation window $T_a$ is typically limited to the eddy turnover time of the fastest measured shell. This poses a challenge for assimilating slower (i.e., lower-wavenumber) shells, which require longer windows. To address this, the 4D-Var process is performed sequentially: a long assimilation interval is divided into smaller windows, and background information is passed from one window to the next. A widely used strategy in numerical weather prediction to perform such sequential assimilation is the ensemble-based 4D-Var (En4D-Var)~\cite{ensemble_atmosperic}. In this approach, multiple 4D-Var optimizations $\tilde{\bm{u}}^{(j)}(t)$ (with $j = 1, \dots, L$ denoting the ensemble index) are run in parallel. Each member uses the background information $\hat{\bm{u}}^{(j)}(0)$, which comes from the posterior solution of the previous sequence, and randomly perturbed observations, the random perturbations are drawn from the same Gaussian distribution as the observation error itself. The background-error covariance matrix $\hat{\bm{\Sigma}}_{\mathbb{C}} = \langle(\hat{\bm{u}} - \langle \hat{\bm{u}} \rangle_L)(\hat{\bm{u}} - \langle \hat{\bm{u}} \rangle_L)^\dagger\rangle_L$ propagates the flow-dependent information across assimilation cycles. The modified cost function used in this ensemble-sequential version is:
\begin{equation}
\label{Jij_seq}
\begin{aligned}
J^{(j)}_{n_{\text{seq}}} =
&\left\| \hat{\bm{u}}^{(j)}(n_{\text{seq}}T_a) - \tilde{\bm{u}}^{(j)}(n_{\text{seq}}T_a) \right\|_{\hat{\bm{\Sigma}}_{\mathbb{C}}} \\
+\sum_{t = n_{\text{seq}}T_a}^{(n_{\text{seq}}+1)T_a}&\left\| \bm{z}(t) - \bm{H}_c \tilde{\bm{u}}^{(j)}(t) - \bm{v}^{(j)}(t) \right\|_{\bm{R}_\mathbb{C}}\,,
\end{aligned}
\end{equation}
where $n_{\text{seq}}$ refers to the sequence index, ranging from 0 to $N_{\text{seq}}$, ensuring that both EnKF and En4D-Var cover the same total assimilation time.

For the first sequence ($n_{\text{seq}}=0$), an ensemble $\hat{\bm{u}}^{(j)}$ is initialized as in \hyperref[eqn:initial_guess_shell_condition_en4dvar]{Eq.~(\ref*{eqn:initial_guess_shell_condition_en4dvar})} below and 4D-Var calculations are solved for a maximum of 50 iterations or until the reduction in the cost function reaches a tolerance. For later sequences ($n_{seq} > 0$), the initial guess is the same as the background information, which is available from the previous sequence as $\hat{\bm{u}}^{(j)}(n_{seq}T_{a}) = \tilde{\bm{u}}^{(j)}(n_{seq}T_{a})$. We also note that the first sequence does not have background information provided.

Before proceeding with the comparison to the EnKF, we briefly summarize some technical details relevant to the application of En-4DVar. First of all, regarding ensemble initialization, we employed a strategy that differs from the one used for EnKF, but is equally optimized. While such tailored initialization would be unfeasible in operational settings, in this theoretical framework we aim to compare the methods under conditions that best highlight the strengths of each approach, including carefully chosen initial states. For En4D-Var, which relies on variational optimization, the convergence speed is sensitive to the initial spread of the ensemble. We therefore seek a strategy that balances ensemble diversity with informativeness. On the one hand, a larger initial spread allows for broader exploration of the state space; on the other hand, it may hinder convergence in variational frameworks like En4D-Var, where optimization performance depends on how closely the ensemble approximates the true state. This highlights a trade-off between convergence efficiency and the risk of ensemble collapse. In En4D-Var experiments, the initial state of each ensemble member \( \hat{u}_n^{(j)}(t_0) \), with \( j = 1, \dots, L \) indexing the \( L \) ensemble members, is defined as:
\begin{equation}
\label{eqn:initial_guess_shell_condition_en4dvar}
\hat{u}_n^{(j)}(t_0)=
\begin{cases}
\sqrt{\langle |u_n|^2 \rangle_T} \left[ \cos(\phi_n^{(j)}) + i \sin(\phi_n^{(j)}) \right] & n \neq m \\
u_n(t_0) + \epsilon_n^{(j)} & n = m\
\end{cases}\ ,
\end{equation}
where \( m \) denotes the observed shells, \( \phi_n^{(j)} \in [0, 2\pi] \) are random phases, and $\epsilon_n^{(j)}$ zero-mean Gaussian noise with covariance $\bm{R}_\mathbb{C}$. In this setup, observed components are initialized using available measurements, while unobserved components are assigned randomized phases scaled by the long-term average energy, ensuring consistency with the system’s statistical structure.

Regarding the execution of the algorithm, a regularization matrix $\mathcal{P}$ is added to the Hessian, where $\mathcal{P}$ is a diagonal matrix defined as $\mathcal{P}(n,n) = N_{\text{reg}}$ for $n < 15$ and $\mathcal{P}(n,n) = N_{\text{reg}} / (2 E_n)$ otherwise. This matrix does not contribute to the cost function or its gradient, but it helps to regularize the Hessian inversion.

In En4D-Var, which operates in complex space, the empirical covariance $\hat{\Sigma}_{\mathbb{C}}$ does not encode phase correlations. Instead, these correlations are captured in EnKF, which works in real space. This difference may lead to minor discrepancies between the two methods, but is not expected to affect the main conclusions of the study.

We consider two outputs from En4D-Var: the \emph{smoother}, which corresponds to the state at the beginning of the window and uses future observations, and the \emph{filter}, which corresponds to the forecast at the end of the window and uses only past data. The filter is directly comparable to the EnKF analysis.  
\hyperref[fig:4dvar_enfk]{Figure~\ref*{fig:4dvar_enfk}} compares EnKF and En4D-Var in terms of (a) energy spectra,  (b) normalized synchronization error and (c) normalized flux-based error.

The EnKF is run with $L = 1000$ ensemble members, while En4D-Var uses $L = 40$ for the $n = 1,2,3$ case, and $L = 160$–$320$ for the $n = 6,7,8$ case. The assimilation window for each sequence is $T_a = 1000\,dt \approx 0.02\,\tau_0 \approx \tau_9$. Similarly to EnKF, inflation could also be used in En4D-Var calculations. This may help to achieve the predictions with fewer ensemble members, but our results with inflation were inconclusive. This is because inflation is empirical and requires fine-tuning the inflation parameter, which is computationally expensive for En4D-Var calculations. Therefore, we do not use inflation for En4D-Var calculations.

The results show that both methods perform similarly, particularly when compared to the filter component of En4D-Var. The main discrepancies appear in the flux error for the smoother in the $n=6,7,8$ case (\hyperref[fig:4dvar_enfk]{Fig.~\ref*{fig:4dvar_enfk}(c)}). These could potentially be reduced by increasing the ensemble size or decreasing the assimilation window $T_a$.

In summary, ensemble-based methods are necessary to assimilate scales beyond those directly measured. Both EnKF and En4D-Var perform well when three consecutive shells are observed. However, the current implementation of En4D-Var is significantly more computationally expensive than the stochastic EnKF method, which is parallelisable in both the forecast and ensemble steps. A more practical implementation of En4D-Var could employ a low-resolution model and quasi-Newton optimization in the inner loop. Such a study on trade-off between accuracy and computational cost is beyond the scope of the present study. Furthermore, 4D-Var also requires constructing the adjoint model, which can be challenging in complex flow systems.
\raggedbottom
%\pagebreak
%%%%%%%%%%%%% End Section Comparison with 4D-Var %%%%%%%%%%%%%
%%%%%%%%%%%%%%%%%%%%%%%%%%%%%%%%%%%%%%%%%%%%%%%%%%%%%%%%%%%%%%

%%%%%%%%%%%%% Section Conclusion %%%%%%%%%%%%%
%%%%%%%%%%%%%%%%%%%%%%%%%%%%%%%%%%%%%%%%%%%%%%
\section{Conclusions}
\label{sec:conclusion}
In this study, we systematically investigated the potential of EnKF for data assimilation of a multiscale turbulent model. Furthermore, some performance against two other DA benchmarks based on En4D-Var and Nudging have been presented. 

We domostrated that the EnKF, when combined with a scale-aware inflation strategy, achieves near-complete synchronization across a wide range of scales, including those not directly observed, provided that measurements are performed at a frequency higher than the characteristic turnover time of the observed shells. Moreover, highly intermittent and non-Gaussian high-frequency scales are also typically reconstructed better than using a statistical baseline. In particular, we found that measuring two adjacent shells in the inertial range is sufficient to reconstruct the full dynamics of the slower scales, while isolated measurements fail to sustain a coherent energy cascade and can destabilize the whole assimilation. Comparisons with Nudging highlighted the limitations in reconstructing unobserved scales due to its strictly localized corrections, while En4D-Var performed comparably to EnKF but at a much higher computational cost.

Taken together, our results indicate that the choice of measurement scales, observation frequency, and ensemble-based correction strategies critically impacts the success of DA in chaotic multiscale systems. They also provide a practical set of guidelines for designing assimilation schemes that balance stability, accuracy, and computational efficiency for realistic applications to Navier-Stokes turbulence in 2d and 3d. 
Future work will extend this framework to more physically realistic systems, such as Rayleigh–Bénard convection flows, where the performance of EnKF could be directly compared against classical Nudging~\cite{agasthya_rayleigh_2022} methods and more recent machine-learning-based approaches like Physics-Informed Neural Networks (PINNs)~\cite{di_leoni_pinns_rayleigh_2023}. Such studies would provide valuable information on the relative strengths of classical and modern DA strategies in fully turbulent settings.\\
Another promising direction involves integrating inflation optimization into a machine learning training framework. Since inflation effectively mimics the effect of larger ensemble sizes, learning the optimal inflation coefficients would allow us to simulate the behavior of massive ensembles without incurring their computational cost. A possible strategy is to adopt a solver-in-the-loop approach~\cite{solver_in_the_loop,freitas2025}, where inflation parameters are updated through backpropagation throughout the assimilation time window, rather than in a single analysis step. This would enable a global optimization of the assimilation process and open new pathways for hybrid dynamical–ML data assimilation methods~\cite{cheng_ml_da_uq_2023}.
%%%%%%%%%%%%% End Conclusion %%%%%%%%%%%%%
%%%%%%%%%%%%%%%%%%%%%%%%%%%%%%%%%%%%%%%%%%

%%%%%%%%%%%%%%%%%%%%%%%%%%%%%%%%%%%%%%%%%%%%%%%%%%%%%%%%%%%%%%%%%%%%%%%%%%%%%%
%%%%%%%%%%%%%%%%%%%%%%%%%%%%%%%%%%%%%%%%%%%%%%%%%%%%%%%%%%%%%%%%%%%%%%%%%%%%%%
\begin{table*}[]
  \label{tab:lambda}
  \centering
  %----------- Left Table (a) ------------
  \begin{minipage}{0.48\textwidth}
    \centering
    %\captionof*{table}{(a) Measurement Sparseness.}
    \vspace{1ex}
    \noindent{(a) Measurement Sparseness.}
    \vspace{0.5ex}
    \begin{ruledtabular}
    \begin{tabular}{ccc}
      6, 7, 8 & $\lambda$ & $N_{\text{inflation}}$ \\
      \hline\specialrule{0.001pt}{0pt}{0pt}
      $\Delta t_{\text{obs}}=\tau_{15}=0.002\tau_0$ & $\lambda=0.2$ & $1$\\
      \specialrule{0.001pt}{0pt}{0pt}
      $\Delta t_{\text{obs}}=\tau_{9}=0.02\tau_0$   & $\lambda=0.1$ & $1$\\
      \specialrule{0.001pt}{0pt}{0pt}
      $\Delta t_{\text{obs}}=\tau_{8}=0.04\tau_0$ &
      $0.01\leq\lambda\leq0.05$ & $6$\\
      \specialrule{0.001pt}{0pt}{0pt}
      $\Delta t_{\text{obs}}=\tau_{7}=0.06\tau_0$ &
      $0.005\leq\lambda\leq0.05$ & $8$\\
      \specialrule{0.001pt}{0pt}{0pt}
      $\Delta t_{\text{obs}}=\tau_{6}=0.1\tau_0$    & $\lambda=0$ & $0$\\
      \specialrule{0.001pt}{0pt}{0pt}
      $\Delta t_{\text{obs}}=\tau_{4}=0.2\tau_0$    & $\lambda=0$ & 0\\
    \end{tabular}
    \end{ruledtabular}
  \end{minipage}
  \hfill
  %----------- Right Table (b) ------------
  \begin{minipage}{0.48\textwidth}
    \centering
    %\captionof*{table}{(b) Measured Scale Schemes.}
    \vspace{1ex}
    \noindent{(b) Measured Scale Schemes.}
    \vspace{0.5ex}
    \begin{ruledtabular}
    \begin{tabular}{ccc}
      $\Delta t_{\text{obs}}=\tau_{15}=0.002\tau_0$ & $\lambda$ & $N_{\text{inflation}}$ \\
      \hline\specialrule{0.001pt}{0pt}{0pt}
      %6, 14         & $\lambda=0$ & $0$\\
      %\specialrule{0.001pt}{0pt}{0pt}
      %4, 14         & $\lambda=0$ & $0$\\
      %\specialrule{0.001pt}{0pt}{0pt}
      %2, 14         & $\lambda=0$ & $0$\\
      %\specialrule{0.001pt}{0pt}{0pt}
      6             & $\lambda=0$ & $0$\\
      \specialrule{0.001pt}{0pt}{0pt}
      6, 7          & $0.1\leq\lambda\leq0.25$ & $8$\\
      \specialrule{0.001pt}{0pt}{0pt}
      6, 7, 8       & $\lambda=0.2$ & $1$\\
      \specialrule{0.001pt}{0pt}{0pt}
      6, 9, 10      & $\lambda=0.1$ & $2$\\
      \specialrule{0.001pt}{0pt}{0pt}
      6, 11, 12     & $0.2\leq\lambda\leq0.25$ & $3$\\
    \end{tabular}
    \end{ruledtabular}
  \end{minipage}
  \caption{Range of inflation strength coefficients $\lambda$ used in both experiments:(a) varying the observation sparseness $\Delta t_{\text{obs}}$ while measuring shells 6, 7, and 8;(b) fixing $\Delta t_{\text{obs}}$ and varying the set of measured shells. In the last column number of experiments (out of 16) that need for inflation ($\lambda = 0$ indicates that no inflation is required).}
\end{table*}
%%%%%%%%%%%%%%%%%%%%%%%%%%%%%%%%%%%%%%%%%%%%%%%%%%%%%%%%%%%%%%%%%%%%%%%%%%%%%%
%%%%%%%%%%%%%%%%%%%%%%%%%%%%%%%%%%%%%%%%%%%%%%%%%%%%%%%%%%%%%%%%%%%%%%%%%%%%%%

%%%%%%%%%%%%% Acknowledgments %%%%%%%%%%%%%
%%%%%%%%%%%%%%%%%%%%%%%%%%%%%%%%%%%%%%%%%%%
\begin{acknowledgments} 
The authors acknowledge useful discussions with Mauro Sbragaglia (University of Rome "Tor Vergata") and Minping Wan (Southern University of Science and Technology, SUSTech).

This research is supported by European Union’s HORIZON MSCA Doctoral Networks programme under Grant Agreement No. 101072344 project AQTIVATE (Advanced computing, QuanTum algorIthms and
data-driVen Approaches for science, Technology and Engineering), by the European Research Council (ERC) under the European Union’s Horizon 2020 research and innovation programme Smart-TURB (Grant Agreement No. 882340), and  the MUR-FARE project R2045J8XAW. AC has been funded as members of the Scale-Aware Sea Ice Project (SASIP) supported by grant G-24-66154 of Schmidt Sciences, LLC –– a philanthropy that propels scientific knowledge and breakthroughs towards a thriving world.
\end{acknowledgments}
%%%%%%%%%%%%% End Acknowledgments %%%%%%%%%%%%%
%%%%%%%%%%%%%%%%%%%%%%%%%%%%%%%%%%%%%%%%%%%%%%%

%%%%%%%%%%%%% Appendices %%%%%%%%%%%%%
%%%%%%%%%%%%%%%%%%%%%%%%%%%%%%%%%%%%%%
\section{Appendices}
%%%%%%%%%%%%% App A : Exponential RK4 %%%%%%%%%%%%%
%%%%%%%%%%%%%%%%%%%%%%%%%%%%%%%%%%%%%%%%%%%%%%%%%%%
\appendix
\section{Runge-Kutta with exponential factor}
\label{app:rk4}
Let $\bm{u} = \{u_0, u_1, \ldots, u_{N-1}\} \in \mathbb{C}^N$ be the system variables, evolving according to \hyperref[eqn:shell_model]{Eq.~(\ref*{eqn:shell_model})}. It is possible to exactly integrate the linear dissipative term $\nu k_n^2 u_n$ by defining the new variable $y_n$ as follows:
\begin{equation}
\label{eqn:factor_appendix}
u_n(t) = y_n(t) I_n(t)\ ,
\end{equation}
where $I_n(t)$ is the \textit{integrating factor}. Making this definition explicit in \hyperref[eqn:shell_model]{Eq.~(\ref*{eqn:shell_model})}, we obtain:
\begin{equation}
\label{eqn:factor_explanation_appendix}
    I_n\frac{dy_n}{dt} + y_n\frac{dI_n}{dt} + \nu k_n^2 I_n y_n = G_n[\bm{I} \circ \bm{y}] + f_n(t)\ .
\end{equation}
Here, the $\circ$ operator denotes the Hadamard product, which performs element-wise multiplication of two vectors, in this case $\bm{I}$ and $\bm{y}$. By imposing that the linear part disappears in terms of the new variable $y_n$:
\begin{equation}
\label{eqn:new_shell_model_appendix}
    y_n\frac{dI_n}{dt} + \nu k_n^2 I_n y_n = 0\ ,
\end{equation}
we find that the integrating factor takes the form:
\begin{equation}
    I_n(t) = \exp\left(-\int dt\, \nu k_n^2\right) = \exp(-\nu k_n^2 t)\ . 
\end{equation}

Considering this expression, the linear term disappears from the evolution equation in terms of the new variable:
\begin{equation}
\frac{dy_n(t)}{dt} = (G_n[I \circ y] + f_n) I_n^{-1}(t)\ .
\label{eqn:new}
\end{equation}
Instead of integrating this last equation, recalling that $u_n(t) = y_n(t) I_n(t)$, we can apply the Runge-Kutta algorithm directly to the original variable $u_n$, obtaining (with a little algebra):
\begin{equation}
\label{eqn:rk4_scheme}
u_n(t + dt) = I_n^2(dt) \left(u_n(t) + \frac{A_{1n}}{6}\right) + I_n \frac{(A_{2n} + A_{3n})}{3} + \frac{A_{4n}}{6}\ ,
\end{equation}
where $I_n(dt) = \exp(-\nu\, dt\, k_n^2 /2)$. Regarding the increments, these are given by:
\begin{equation}
\begin{cases}
A_{1n} = dt(f_n + G_n[\bm{u}]) \\
A_{2n} = dt(f_n + G_n[\bm{I} \circ (\bm{u} + \frac{\bm{A}_1}{2})]) \\
A_{3n} = dt(f_n + G_n[\bm{I} \circ \bm{u} + \frac{\bm{A}_2}{2}]) \\
A_{4n} = dt(f_n + G_n[\bm{I} \circ \bm{I} \circ \bm{u} + \bm{I} \circ \bm{A}_3])
\end{cases}
\end{equation}
where $\bm{I} = \{e^{-\nu dt k_0^2/2}, e^{-\nu dt k_1^2/2}, \ldots, e^{-\nu dt k_{N-1}^2/2}\}$.

Finally, nudging is treated using the same technique, namely by integrating the linear terms in \hyperref[eqn:nudging]{Eq.~(\ref*{eqn:nudging})} through the introduction of appropriate integrating factors:
\begin{equation}
    I_{nm} = e^{-dt (\nu k_n^2 + \alpha_n \delta_{nm})/2}\ ,
\end{equation}
where $m$ indicates the measured shell. The state is evolved by using the same as in \hyperref[eqn:rk4_scheme]{Eq.~(\ref*{eqn:rk4_scheme})} but with coefficient
\begin{equation}
\begin{cases}
A_{1n} = dt(f_n + G_n[\bm{u}]+T_n(\bm{z},t)) \\
A_{2n} = dt(f_n + G_n[\bm{I} \circ (\bm{u} + \frac{\bm{A}_1}{2})]+T_n(\bm{z},t+\frac{dt}{2})) \\
A_{3n} = dt(f_n + G_n[\bm{I} \circ \bm{u} + \frac{\bm{A}_2}{2}]+T_n(\bm{z},t+\frac{dt}{2})) \\
A_{4n} = dt(f_n + G_n[\bm{I} \circ \bm{I} \circ \bm{u} + \bm{I} \circ \bm{A}_3]+T_n(\bm{z},t+dt))
\end{cases}
\end{equation}
where \[
\begin{aligned}
\bm{I} = \{&e^{-dt(\nu k_0^2+\delta_{0m}\alpha_0)/2},
            e^{-dt(\nu k_1^2+\delta_{1m}\alpha_1)/2}, \ldots, \\
          &e^{-dt(\nu k_{N-1}^2+\delta_{N-1,m}\alpha_{N-1})/2}\}\ .
\end{aligned}
\]

%%%%%%%%%%%%% End App A : Exponential RK4 %%%%%%%%%%%%%
%%%%%%%%%%%%%%%%%%%%%%%%%%%%%%%%%%%%%%%%%%%%%%%%%%%%

%%%%%%%%%%%%%%%%%%%%%%%%%%%%%%%%%%%%%%%%%%%%%%%%%%%%%%%%%%%%%%%%%%%%%%%%%%%%%%
%%%%%%%%%%%%%%%%%%%%%%%%%%%%%%%%%%%%%%%%%%%%%%%%%%%%%%%%%%%%%%%%%%%%%%%%%%%%%%
\section{Numerical Details on Inflation}
\label{app:inflation}
For completeness, \hyperref[tab:lambda]{Table~\ref{tab:lambda}} provides a detailed summary of the inflation parameters used throughout the experiments. These parameters were chosen to stabilize the assimilation process while simultaneously achieving the lowest possible MSE.

%\nocite{*}
%\bibliographystyle{apsrev4-2}
%\bibliography{main}% Produces the bibliography via BibTeX.

\bibliographystyle{apsrev4-2}
\providecommand{\noopsort}[1]{}\providecommand{\singleletter}[1]{#1}%

\vspace{0.1cm}
\hypertarget{emailnote}{\textsuperscript{\textcolor{blue}{*}}\href{mailto:francesco.fossella@telecom-paris.fr}{francesco.fossella@telecom-paris.fr}}

\end{document}